%% file: Project_1_Masterfile.tex
\documentclass[a4paper,leqno]{article}
\usepackage[latin1]{inputenc}
\usepackage[T1]{fontenc}	
\usepackage{amsmath,amssymb,amsfonts}
\usepackage{accents}
\usepackage[titletoc]{appendix}
\newcommand{\interior}[1]{\accentset{\smash{\raisebox{-0.12ex}{$\scriptstyle\circ$}}}{#1}\rule{0pt}{2.3ex}}
\usepackage[margin=3cm]{geometry}
\usepackage{graphicx}
\graphicspath{{./used_plots/}} 
\usepackage{subfigure}
\usepackage[format=plain,labelsep=quad,isu]{caption}
\newtheorem{remark}{Remark}
\def \asinh {\text{asinh}}

\def \ln {\text{ln}}

\begin{document}
\title{High-order compact finite difference schemes for option pricing
  in stochastic volatility models on non-uniform grids}
\author{Bertram D{\"u}ring \thanks{Email:~b.during@sussex.ac.uk, Department of Mathematics, University of Sussex, Pevensey II, Brighton, BN1 9QH, United Kingdom}\and Michel Fourni\'e
\thanks{Email:~michel.fournie@math.univ-toulouse.fr, Institut de
  Math\'ematiques de Toulouse, Universit\'e de Toulouse et CNRS (UMR 5219), France}
\and Christof Heuer
\thanks{Email:~c.heuer@sussex.ac.uk, Department of Mathematics, University of Sussex, Pevensey II, Brighton, BN1 9QH, United Kingdom}} 
\maketitle
\begin{abstract}
\noindent We derive high-order compact finite difference schemes for option
pricing in stochastic volatility models on non-uniform grids. 
The schemes are fourth-order
accurate in space and second-order accurate in time for vanishing
correlation. In our numerical study we obtain high-order numerical
convergence also for non-zero correlation and non-smooth payoffs which
are typical in option pricing. 
In all numerical experiments a comparative standard second-order discretisation is
significantly outperformed.
We conduct a numerical stability study which indicates
unconditional stability of the scheme.
\end{abstract}


\input{intro}
\input{SV_Model_with_zoom}
\begin{appendices}
\input{HestonVersion2_as_Appendix}
\input{HestonVersion4_as_Appendix}
\end{appendices}

\input{biblio.tex}
\end{document}

%% file: intro.tex
\section{Introduction}

Efficient pricing of financial derivatives, in particular options, is one of the major topics in
financial mathematics.
To be able to explain important effects which are present in real financial markets,
e.g.\ the volatility smile (or skew) in option prices, so-called {\em stochastic
volatility\/} models have been introduced over the last two decades.
In contrast to the seminal paper of Black and Scholes \cite{BS73} the underlying
asset's volatility is not assumed to be constant, but is itself
modelled by a stochastic diffusion process.
These stochastic volatility models are typically based on a two-dimensional
stochastic diffusion process with
two Brownian motions with correlation $\rho$, i.e.
$dW^{(1)}(t)dW^{(2)}(t)=\rho\, dt$. On a given filtered probability space for
the stock price $S=S(t)$ and the stochastic volatility $\sigma=\sigma(t)$ one considers
\begin{align*}
dS(t)& =\bar{\mu} S(t)\,dt +\sqrt{\sigma(t)} S(t)\,dW^{(1)}(t),\\
d\sigma(t)& =a(\sigma(t)) \,dt+b(\sigma(t))\,dW^{(2)}(t),
\end{align*}
where $\bar{\mu}$ is the drift of the stock, $a(\sigma)$ and $b(\sigma)$ are the drift and the diffusion coefficient of
the stochastic volatility.

Application of It\^o's Lemma and standard arbitrage arguments show that any
option price $V=V(S,\sigma,t)$ solves the following partial differential equation,
\begin{equation}
\label{P0}
 V_t+\frac12 \sigma S^2\sigma V_{SS}+\rho b(\sigma) \sqrt{\sigma}S V_{S\sigma}+\frac12  b^2(\sigma)
V_{\sigma\sigma}+\bigl(a(\sigma) -\lambda(S,\sigma,t)\bigr) V_\sigma+rSV_S-rV=0, 
\end{equation}
where $r$ is the (constant) riskless interest
rate and $\lambda(S,\sigma,t)$ denotes the market price of volatility
risk.
Equation \eqref{P0} has to be solved for
$S,\sigma>0,\;0 \leq t \leq T,$ and subject to final and 
boundary conditions which depend on the specific option that is to be
priced. 

There are different stochastic volatility models with different
choices of the model for the evolution of the volatility for $t>0$, starting
from an initial volatility $\sigma(0)>0.$
 The most prominent work in this direction is the Heston model \cite{Heston}, where 
\begin{align}
 \label{eq:HestonVola}
 d\sigma(t)& =\kappa^*\bigl(\theta^*-\sigma(t)\bigr) \,dt+v\sqrt{\sigma(t)}\,dW^{(2)}(t).
 \end{align}
Other stochastic volatility models are, e.g., the GARCH diffusion model \cite{Duan95},
\begin{equation}
\label{eq:garchmodel}
d\sigma(t)=\kappa^*\bigl (\theta^*-\sigma(t) \bigr)\,dt+v\sigma(t)\,dW^{(2)}(t),
\end{equation}
or the so-called 3/2-model (see, e.g.\ \cite{Lewis00}),
\begin{equation}
\label{eq:32model}
d\sigma(t)= \kappa^*\sigma(t)\bigl(\theta^*-\sigma(t) \bigr)\,dt+v{\sigma(t)}^{3/2}\,dW^{(2)}(t).
\end{equation}
In \eqref{eq:HestonVola}-\eqref{eq:32model}, $\kappa^*$, $v$, and
$\theta^*$ denote the mean reversion speed, the volatility of volatility, and the
 long-run mean of $\sigma,$ respectively.

For some models and under additional restrictions, closed form
solutions to \eqref{P0} can be obtained by Fourier methods (see, e.g.\
  \cite{Heston,Due09}). 
Another approach is to derive approximate analytic expressions, see, e.g.\
\cite{BeGoMi10} and the literature cited therein.
In general, however, ---even in the Heston model when the
parameters are non constant--- equation \eqref{P0} has
to be solved numerically. Moreover, many (so-called American) options
feature an additional early exercise right. Then one has to solve a
free boundary problem which consists of \eqref{P0} and an early
exercise constraint for the option price. Also for this problem one
typically has to resort to numerical approximations.

In the mathematical literature, there are a number of papers
considering numerical methods for option
pricing in stochastic volatility models, i.e.\ for two spatial 
dimensions. Finite difference approaches that are used are
often standard, low order methods (second order in space). Other
approaches include finite element-finite
volume \cite{ZvFoVe98}, multigrid \cite{ClaPar99}, sparse wavelet
\cite{HiMaSc05}, or spectral methods \cite{ZhuKop10}.

Let us review some of the related finite difference literature. Different efficient
methods for solving the American
option pricing problem for the Heston model are compared in
\cite{IkoToi07}. The article focusses on the treatment of the early
exercise free boundary 
and uses a second order finite difference discretization.
In \cite{HouFou07} different, low order ADI (alternating
direction implicit) schemes are adapted to the Heston model to include
the mixed spatial derivative term. 
While most of \cite{TaGoBh08} focusses on high-order compact scheme
for the standard (one-dimensional) case, in a short remark \cite[Section~5]{TaGoBh08} also the
stochastic volatility (two-dimensional) case is considered. However,
the final scheme is of second order only due to the low order
approximation of the cross diffusion term.

High-order finite difference schemes (fourth order in space)
were proposed for option pricing with deterministic (or constant) volatility,
i.e. in one spatial dimension, that use a compact stencil (three
points in space), see, e.g., \cite{TaGoBh08} for linear and
\cite{DuFoJu03,DuFoJu04,LiaKha09} for fully nonlinear problems. 

More recently, a high-order compact finite difference scheme for
(two-dimensional) option pricing 
models with {\em stochastic volatility\/} has been presented in \cite{DuFo12}. 
This scheme uses a uniform mesh and is fourth order accurate in space
and second order accurate in time. Unconditional (von Neumann)
stability of the scheme is proved for vanishing correlation. A
further study of its stability, indicating unconditional stability
also for non-zero correlation, is performed in \cite{DuFo12p}.

In general, the accuracy of a numerical discretisation of \eqref{P0} for a given
number of grid points can be greatly improved by considering a
{\em non-uniform mesh}. This is particular true for option pricing
problems as \eqref{P0},  as typical initial conditions have a discontinuity
in their first derivative at $S=K$, which is the center of the area of interest (`at-the-money').

Our aim in the present paper is to consider extensions of the
high-order compact methodology for stochastic volatility models 
\eqref{P0} to non-uniform grids.
The basic idea of our approach is to introduce a transformation of the
partial differential equation from a non-uniform grid to a uniform
grid (as, e.g.\ in \cite{Fournie00}). 
Then, the high-order compact methodology can be applied to this
transformed partial differential equation. It turns out, however, that
this process is not straight-forward as the derivatives of the transformation appear in
the truncation error and due to the presence of the cross-derivative 
terms, one cannot proceed to cancel terms in the truncation error in a
similar fashion as in \cite{DuFo12} and the derivation of a
high-order compact scheme becomes much more involved. 
Nonetheless, we are able to derive a compact scheme which shows
high-order convergence for typical European option pricing problems.
Up to the knowledge of the authors, this is the first high-order
compact scheme for option pricing in stochastic volatility models on
non-uniform grids.

The rest of this paper is organised as follows. In the next section,
we transform \eqref{P0} into a more convenient form. We then derive
four new variants of a compact scheme in Section~\ref{sec:derivation}.
Numerical experiments confirming the high-order convergence for
different initial conditions (we consider the case of a European Put
option and a European Power Put option) are presented in
Section~\ref{Application_of_Version_3_Heston_with_zoom}. Section~\ref{sec:conc} concludes.


%% file: SV_Model_with_zoom.tex

\section{Transformation of the partial differential equation and final condition}
\label{sec:trans}

We focus our attention on the Heston model \eqref{P0}--\eqref{eq:HestonVola}, although our methodology
adapts also to other stochastic volatility models in a natural way
(see Remark~\ref{remark} at the end of Section~\ref{sec:derivation}).
As usual, we restrict ourselves to the case where
the market price of volatility risk $\lambda(S,\sigma,t)$ is
proportional to $\sigma$ and choose
$\lambda(S,\sigma,t)=\lambda_0\sigma$ for some 
 constant $\lambda_0$.
This allows to study the problem using the modified parameters
 $$
 \kappa=\kappa^*+\lambda_0,\quad \theta=\frac{\kappa^*\theta^*}{\kappa^*+\lambda_0},
 $$
 which is both convenient and standard practice. For similar reasons,
 some authors set the market price of volatility risk to zero.

The partial differential equation of the Heston-model is then given by
\begin{equation}\label{Heston_model_original_pde}
V_t+\frac12 \sigma S^2\sigma V_{SS}+\rho v \sigma S V_{S\sigma}+\frac12 v^2 \sigma
V_{\sigma\sigma}+rSV_S+ \kappa(\theta - \sigma)  V_\sigma-rV=0
\end{equation}
where $S\in \bigl[0,S_{\max}\bigr]$ with a chosen $S_{\max}>0$,
$\sigma\in \left[\sigma_{\min},\sigma_{\max}\right]$ with $0\leq
\sigma_{\min}<\sigma_{\max}$ and $t \in \left[0,T\right[$ with $T>0$,
imposing an approximative artificial boundary condition at $S_{\max}$. 
The error caused by approximative boundary conditions imposed on an artificial
boundary for a class of Black-Scholes equations has been studied
rigorously in \cite{KaNi00}. 

The final condition as well as the boundary conditions, which we will
discuss separately, depend on the chosen option. In the case of a
European Power Put Option we have the final condition
\begin{eqnarray}\label{Final_Condition_Original_Problem}
V(S,v,T)=\max(K-S,0)^p
\end{eqnarray}
with power $p\in \mathbb{N}$.

For high-order finite difference schemes as proposed in this article,
the low regularity of the final
condition \eqref{Final_Condition_Original_Problem} 
at the strike $S=K$ may
reduce the numerical convergence order in practice. To retain
high-order convergence, one can smooth the initial condition
carefully (cf.\ \cite{KrThWi70}) or shift the numerical grid to avoid the strike
falling on a grid point as suggested, for example, in
\cite{Tavella,DuFo12}. In our
numerical experiments reported in
Section~\ref{Application_of_Version_3_Heston_with_zoom} we use the
latter approach.

We apply the following transformations to \eqref{Heston_model_original_pde} as in \cite{DuFo12},
$$
\hat{S}=\ln\left(\frac{S}{K}\right),\quad  \tau = T-t,\quad
y=\frac{\sigma}{v}, \quad  u=e^{r\tau}\frac{V}{K},
$$
where $\hat{S}\in \left[\hat{S}_{\min}, \hat{S}_{\max} \right]$ with a
chosen $\hat{S}_{\min}<0$ and
$$
\hat{S}_{\max}=\ln\left(\frac{S_{\max}}{K}\right).
$$
We then introduce a (sufficiently smooth) zoom function
$$
\hat{S}=\varphi(x),
$$ 
zooming around $\hat{S}=0$, with 
$$x \in \left[\varphi^{-1} \left(\hat{S}_{\min} \right) , \varphi^{-1} \left(\hat{S}_{\max} \right) \right],$$ 
and setting 
$f=-u_{\tau}$ we obtain from \eqref{Heston_model_original_pde} the
following two-dimensional elliptic problem,
\begin{equation}\label{pdezudiskretisierenII}
\varphi_x^3 f = \frac{-vy}{2}\left[\varphi_x u_{xx} + \varphi_x^3 u_{yy}\right] - \rho vy\varphi_x^2 u_{xy} - \kappa \frac{\theta -vy}{v}\varphi_x^3 u_y + \left[\frac{vy\varphi_{xx}}{2} + \Bigl(\frac{vy}{2}-r\Bigr)\varphi_x^2 \right] u_x ,
\end{equation}
where $(x,y) \in \Omega:=[x_{\min},x_{\max}]\times [y_{\min},y_{\max}]$, $x_{\min}<x_{\max}$ and $y_{\min}<y_{\max}$. 

\section{Derivation of the high-order compact schemes for the elliptic problem}
\label{sec:derivation}

We start by defining a uniform grid in $x$- and in $y$-direction,
\begin{eqnarray} \label{Definition_of_general_grid}
 G := \left\{ (x_i,y_j) \in \Omega\; |\; x_i= x_{\min} + i(\Delta x),\;y_i = y_{\min} + j(\Delta y), \;0\leq i \leq N,\; 0 \leq j \leq M \right\},
\end{eqnarray}
where $\Delta x=(x_{\max} - x_{\min})/N$ and $\Delta y=(y_{\max} -
y_{\min})/M$ are the step sizes in each direction. With $\interior{G}$
we identify the inner points of the grid $G$. On this grid we denote
by $U_{ij}$ the discrete approximation of the continuous solution $u$ in $(x_i,y_j) \in G$. Using the standard
central difference operator $D_x^c$ in
$x$-direction and $D_y^c$ in $y$-direction, and the standard second-order
central difference operator $D_x^2$ in $x$-direction and $D_y^2$ in
$y$-direction, for $k=x,y$ we have 
\begin{eqnarray}\label{consistencyequations}
  \begin{array}{rcl}
u_k&=&D^c_k U_{ij} - \frac{(\Delta k)^2}{6}u_{kkk}+ \mathcal{O}\left((\Delta k)^4\right),\\
\end{array}
\end{eqnarray}
and
\begin{eqnarray}\label{consistencyequationsII}
\begin{array}{rcl}
u_{kk}&=&D^2_k U_{ij} - \frac{(\Delta k)^2}{12}u_{kkkk} +
\mathcal{O}\left((\Delta k)^4\right),\\
\\
u_{xy}&=&D^c_xD^c_y U_{ij} - \frac{(\Delta x)^2}{6}u_{xxxy} - \frac{(\Delta y)^2}{6}u_{xyyy}+ \mathcal{O}\left((\Delta x)^4\right)\\
&&  + \mathcal{O}\left((\Delta x)^2 (\Delta y)^2\right) + \mathcal{O}\left((\Delta y)^4\right)+ \mathcal{O}\left(\frac{(\Delta x)^6}{\Delta y}\right),
\end{array}
\end{eqnarray}
at the grid points $(x_i,y_j)$ for $i = 0, \ldots, N $ and $j = 0 , \ldots, M$. We call a scheme of high order, if its consistency error is of order
$\mathcal{O}\left((\Delta x)^4\right)$ for $\Delta
y\in\mathcal{O}\left(\Delta x\right)$.
If we discretise the higher derivatives $u_{xxxx}$,
$u_{yyyy}$, $u_{xxxy}$, $u_{xyyy}$, $u_{xxx}$, and $u_{yyy}$ appearing
in \eqref{consistencyequations} and \eqref{consistencyequationsII}
with second order accuracy, we obtain a scheme with
consistency of order four, since they are
all multiplied by factors of order two. If this can be achieved using
the compact nine-point computational stencil,
\begin{eqnarray}
\notag\left( \begin{array}{ccc}
U_{i-1,j+1}&U_{i,j+1}&U_{i+1,j+1}\\
\\
U_{i-1,j}&U_{i,j}&U_{i+1,j}\\
\\
U_{i-1,j-1}&U_{i,j-1}&U_{i+1,j-1}
\end{array} \right),
\end{eqnarray}
the scheme is called \textit{high-order compact (HOC)}. 

\subsection{Auxiliary relations for higher derivatives}\label{sec:Aux_relations}

We proceed by giving auxiliary relations for the third and fourth order derivatives appearing in \eqref{consistencyequations} and \eqref{consistencyequationsII}. Expressions for the higher derivatives can be obtained by differentiating the partial differential equation \eqref{pdezudiskretisierenII} in a formal manner without introducing additional error. Differentiating equation \eqref{pdezudiskretisierenII} with respect to $x$ and then solving for $u_{xxx}$ leads to
\begin{eqnarray}\label{uxxx}
\begin{array}{rcl}
u_{xxx}& = & -\frac{6 \varphi_x \varphi_{xx}}{vy}f - \frac{2\varphi_x^2}{vy}f_x  + \left[  \frac{\varphi_{xxx}}{\varphi_x} + \frac{4 \left( \frac{vy}{2}-r \right)\varphi_{xx}}{vy} \right]u_x + \frac{2\left( \frac{vy}{2}-r\right) \varphi_x}{vy}u_{xx} - \varphi_{x}^2 u_{xyy} \\
\\
&& - 6\kappa \frac{\theta -vy}{v^2y}\varphi_x\varphi_{xx}  u_y - \left[ 4\rho \varphi_{xx} + 2\kappa \frac{\theta -vy}{v^2y}\varphi_x^2\right] u_{xy}- 2\rho \varphi_x u_{xxy} -  3\varphi_{x} \varphi_{xx} u_{yy} \\
\\
& =: & A_{xxx}.
\end{array}
\end{eqnarray}
Using this equation we can calculate a
discretisation of $A_{xxx}$ using only points of the nine-point stencil
with consistency error of order two using the central difference operators.

Differentiating the partial differential equation
\eqref{pdezudiskretisierenII} twice with respect to $x$ and then solving
for $u_{xxxx}$ we have
\begin{eqnarray}\label{uxxxx}
\begin{array}{rcl}
u_{xxxx}&=& \frac{vy\varphi_{xxxx}  + 4 \left( \frac{vy}{2}-r\right)\left[ \varphi_x \varphi_{xxx} + \varphi_{xx}^2\right]}{vy\varphi_{x}} u_x  +\left[ \frac{\varphi_{xxx}}{\varphi_x } + \frac{8\left(\frac{vy}{2}-r \right)\varphi_{xx}}{vy}\right]u_{xx} \\
\\
&& + \left[ \frac{2 \left( \frac{vy}{2}-r\right)\varphi_x}{vy}- \frac{\varphi_{xx}}{\varphi_x}\right] u_{xxx} - 6\varphi_x \varphi_{xx}u_{xyy} - \varphi_{x}^2 u_{xxyy} \\
\\
&&-\frac{6\kappa(\theta - vy)\left[2 \varphi_{xx}^2 + \varphi_x \varphi_{xxx}\right]}{v^2y} u_y - \left[ 4\rho \left( \varphi_{xxx} + \frac{\varphi_{xx}^2}{\varphi_x}\right) + \frac{12 \kappa \left(\theta -vy\right)\varphi_x \varphi_{xx}}{v^2y}\right] u_{xy}  \\
 \\
&& - \left[ 8\rho\varphi_{xx} + \frac{2 \kappa\left(\theta - vy \right) \varphi_x^2}{v^2y}\right] u_{xxy}- 2 \rho \varphi_x u_{xxxy} - \left[ 3 \varphi_x \varphi_{xxx} +6 \varphi_{xx}^2\right] u_{yy} \\
\\
&&  - \frac{12\varphi_{xx}^2 + 6 \varphi_x \varphi_{xxx}}{vy}f - \frac{12\varphi_x \varphi_{xx}}{vy} f_x - \frac{2\varphi_x^2}{vy}f_{xx} =: A_{xxxx} - 2\rho\varphi_x u_{xxxy}. 
\end{array}
\end{eqnarray}
The term $A_{xxxx}$ can be discretised at the order two on the compact stencil if equation \eqref{uxxx} and the central difference operator are used. Solving equation \eqref{uxxxx} for $u_{xxxy}$ we obtain
\begin{eqnarray}\label{uxxxxSolvedForUxxxy}
u_{xxxy} = \frac{1}{2\rho\varphi_x}A_{xxxx} -\frac{1}{2\rho\varphi_x}u_{xxxx}.
\end{eqnarray}
In order to find an equation for $u_{yyy}$ we first differentiate the partial differential equation \eqref{pdezudiskretisierenII} once with respect to $y$ and then solve for $u_{yyy}$, which leads to
\begin{eqnarray}\label{uyyy}
\begin{array}{rclcl}
u_{yyy}&=& -\frac{1}{\varphi_x^2}u_{xxy} - \frac{1}{y \varphi_x^2}u_{xx} - \frac{2\rho}{\varphi_x} u_{xyy} -\frac{2\kappa(\theta -vy)+ v^2}{v^2y}u_{yy}  \\
\\
&& + \frac{2\kappa}{vy}u_y  + \left[ \frac{\varphi_{xx}}{\varphi_x^3}  + \frac{2\left(\frac{vy}{2} -r\right) - 2\rho v}{vy\varphi_x}\right]u_{xy} + \frac{\varphi_{xx} + \varphi_x^2}{y\varphi_x^3}u_x -\frac{2}{vy}f_y&=:& A_{yyy}.
\end{array}
\end{eqnarray}
The term $A_{yyy}$ can be discretised in a compact manner at the order two
using the central difference operators. 

Differentiating equation \eqref{pdezudiskretisierenII} twice with respect to $y$ and then solving for $u_{yyyy}$ leads to
\begin{eqnarray}\label{uyyyy}
\begin{array}{rcl}
u_{yyyy}&=& -\frac{1}{\varphi_x^2}u_{xxyy} - \frac{2}{y\varphi_x^2}u_{xxy} -\left(\frac{2v^2 + 2\kappa(\theta - vy)}{v^2y}\right)u_{yyy} - \frac{2\rho}{\varphi_x}u_{xyyy} + \frac{4\kappa}{vy}u_{yy} \\
\\
&&+ \frac{2\varphi_{xx} + 2\varphi_x^2}{y\varphi_x^3}u_{xy}  + \left(\frac{\varphi_{xx} }{\varphi_x^3} +\frac{2\left( \frac{yv}{2}-r\right)-4\rho v}{yv\varphi_x}\right)u_{xyy} - \frac{2}{vy}f_{yy}\\
\\
&=:&A_{yyyy} - \frac{2\rho}{\varphi_x}u_{xyyy}.
\end{array}
\end{eqnarray}
The term $A_{yyyy}$ can be discretised at the order two on the compact stencil using equation \eqref{uyyy} and the central difference operator. Equation \eqref{uyyyy} is equivalent to
\begin{eqnarray}\label{uyyyySolvedForUxyyy}
u_{xyyy} = \frac{\varphi_x}{2\rho}A_{yyyy} -\frac{\varphi_x}{2\rho} u_{yyyy}.
\end{eqnarray}
Differentiating the partial differential equation \eqref{pdezudiskretisierenII} once with respect to $x$ and once with respect to $y$ and then solving for $u_{xxxy}$ leads to
\begin{eqnarray}\label{uxxxy}
\begin{array}{rcl}
u_{xxxy}  & = & \left[ \frac{\varphi_{xxx}}{y \varphi_x} + \frac{2 \varphi_{xx}}{y}\right] u_x + \frac{\varphi_x}{y}u_{xx} - \frac{1}{y}u_{xxx} - \left[ \frac{6\kappa(\theta -vy)\varphi_x \varphi_{xx}}{v^2y} + \frac{3\varphi_x \varphi_{xx}}{y}\right]u_{yy} \\
\\
&& + \frac{6\kappa\varphi_x\varphi_{xx}}{vy}u_y  - 3\varphi_x \varphi_{xx} u_{yyy} + \left[ \frac{\varphi_{xxx}}{\varphi_x} - \frac{4\rho \varphi_{xx}}{y} + \frac{4\left( \frac{vy}{2}-r\right)\varphi_{xx}}{vy}+ \frac{2\kappa\varphi_x^2}{vy} \right]u_{xy} \\
\\
&& - 2\rho \varphi_x u_{xxyy} - \left[ \frac{2\kappa(\theta -vy)\varphi_x^2}{v^2y} + 4\rho \varphi_{xx} + \frac{\varphi_x^2}{y}\right]u_{xyy} - \varphi_x^2 u_{xyyy}\\
\\
&& + \left[\frac{2\left( \frac{vy}{2} - r \right) \varphi_x}{vy} - \frac{2\rho \varphi_x}{y} \right] u_{xxy} - \frac{6\varphi_x \varphi_{xx}}{vy}f_y - \frac{2\varphi_x^2}{vy}f_{xy} \\
\\
&=:&A_{xxxy} - \varphi_x^2 u_{xyyy}.
\end{array}
\end{eqnarray}
Using the equations \eqref{uxxx} and \eqref{uyyy} as well as the
central difference operators in $x$- and $y$-direction it is possible
to discretise $A_{xxxy}$ at the order two on the compact
stencil. Solving equation \eqref{uxxxy} for $u_{xyyy}$ gives
\begin{eqnarray}\label{uxyyy}
u_{xyyy} = \frac{A_{xxxy}}{\varphi_x^2 }- \frac{1}{\varphi_x^2 }u_{xxxy} =: A_{xyyy} - \frac{1}{\varphi_x^2 }u_{xxxy}.
\end{eqnarray}
Finally, the expression $A_{xyyy}$ can be discretised at the order two on the
compact stencil as well.

\subsection{Derivation of the discrete schemes}\label{sec:derivation_of_schemes}
In order to derive a discrete scheme we employ equations
\eqref{consistencyequations} and \eqref{consistencyequationsII} in the
partial differential equation \eqref{pdezudiskretisierenII}, which gives
\begin{eqnarray}\label{semidiscretepdezudiskretisierenII}
\begin{array}{rcl}
\varphi_x^3 f &=&A_0 + \varepsilon + \frac{vy(\Delta x)^2\varphi_x}{24}u_{xxxx} +\frac{vy(\Delta y)^2\varphi_x^3}{24}u_{yyyy}+\frac{\rho vy(\Delta x)^2\varphi_x^2 }{6}u_{xxxy}\\
\\
&& + \frac{\rho vy(\Delta y)^2\varphi_x^2 }{6}u_{xyyy} +\frac{\kappa(\theta - vy)(\Delta y)^2\varphi_x^3}{6v}u_{yyy} - \frac{\left[ vy\varphi_{xx} + 2\left(\frac{vy}{2}-r\right)\varphi_x^2\right](\Delta x)^2}{12} u_{xxx} ,
\end{array}
\end{eqnarray}
where
\begin{eqnarray}
\notag \begin{array}{rcl}
A_0 &:=& -\frac{vy}{2}\left[\varphi_x D^2_x U_{ij}  +\varphi_x^3 D^2_y U_{ij}\right] - \rho vy\varphi_x^2  D^c_xD^c_y U_{ij} - \kappa \frac{\theta -vy}{v}\varphi_x^3 D^c_yU_{ij}\\
\\
&&+ \left[ \frac{vy\varphi_{xx}}{2} + \left( \frac{vy}{2} - r \right)\varphi_x^2\right] D^c_x U_{ij} 
\end{array}
\end{eqnarray}
and the error-term $\varepsilon \in \mathcal{O}\left( (\Delta
  x)^4\right)$ if  $\Delta y\in\mathcal{O}\left(\Delta x\right) $ is
used. Equation \eqref{semidiscretepdezudiskretisierenII} is the basis
for the derivation of our different discretisation schemes. $A_0$ is only using the compact stencil. 

We have four fourth-order derivatives, namely
$u_{xxxx}$, $u_{yyyy}$, $u_{xxxy}$ and $u_{xyyy}$ appearing in
equation \eqref{semidiscretepdezudiskretisierenII}, interacting with
each other, but only three auxiliary relations to replace these higher derivatives. These relations are given by \eqref{uxxxx}, \eqref{uyyyy}, and \eqref{uxxxy}, which were derived in Section~\ref{sec:Aux_relations}. 
This leads to four different versions of the discrete scheme. 

For the \textit{Version 1 scheme} equations \eqref{uxxx},
\eqref{uyyy} and \eqref{uyyyy} are used in equation
\eqref{semidiscretepdezudiskretisierenII}, then \eqref{uxyyy} is employed and finally
\eqref{uxxxxSolvedForUxxxy} is applied, which gives 
\begin{eqnarray}\label{Version1}
\begin{array}{rcl}
\varphi_x^3 f&=&A_0 +\frac{ vy\left[2(\Delta x)^2\varphi_x^2 -(\Delta y)^2\right]}{24\varphi_x} A_{xxxx}  +\frac{vy(\Delta y)^2\varphi_x^3}{24}A_{yyyy}  + \frac{\rho vy(\Delta y)^2\varphi_x^2 }{12}A_{xyyy}\\
\\
&&+\frac{\kappa(\theta - vy)(\Delta y)^2\varphi_x^3}{6v}A_{yyy} - \frac{\left[ vy\varphi_{xx} + 2\left(\frac{vy}{2}-r\right)\varphi_x^2\right](\Delta x)^2}{12} A_{xxx}\\
\\
&& + \frac{ vy\left[(\Delta y)^2-(\Delta x)^2\varphi_x^2 \right]}{24\varphi_x} u_{xxxx} + \varepsilon .
\end{array}
\end{eqnarray}
For the \textit{Version 2 scheme} equations \eqref{uxxx}, \eqref{uyyy} and \eqref{uxxxx} are used in equation \eqref{semidiscretepdezudiskretisierenII}, then  \eqref{uxxxy} is employed and finally \eqref{uyyyySolvedForUxyyy} is applied, which gives
\begin{eqnarray}\label{Version2}
\begin{array}{rcl}
\varphi_x^3 f&=&A_0 + \frac{vy(\Delta x)^2\varphi_x}{24}A_{xxxx} + \frac{ vy\varphi_x^3[2(\Delta y)^2 - (\Delta x)^2\varphi_x^2 ]}{24} A_{yyyy}  +\frac{\rho vy(\Delta x)^2\varphi_x^2 }{12}A_{xxxy}\\
\\
&&+\frac{\kappa(\theta - vy)(\Delta y)^2\varphi_x^3}{6v}A_{yyy} - \frac{\left[ vy\varphi_{xx} + 2\left(\frac{vy}{2}-r\right)\varphi_x^2\right](\Delta x)^2}{12} A_{xxx}\\
\\
&&+\frac{vy\varphi_x^3[ (\Delta x)^2\varphi_x^2 -(\Delta y)^2 ]}{24}u_{yyyy}+ \varepsilon .
\end{array}
\end{eqnarray}
For the \textit{Version 3 scheme} equations \eqref{uxxx}, \eqref{uyyy}, \eqref{uxxxx} and \eqref{uyyyy} are used in equation \eqref{semidiscretepdezudiskretisierenII} and then \eqref{uxyyy} is applied, which gives
\begin{eqnarray}\label{Version3}
\begin{array}{rcl}
\varphi_x^3 f &=& A_0 + \frac{vy(\Delta x)^2\varphi_x}{24} A_{xxxx} +\frac{vy(\Delta y)^2\varphi_x^3}{24}A_{yyyy} + \frac{\rho vy(\Delta y)^2\varphi_x^2 }{12}A_{xyyy} \\
\\
&&+\frac{\kappa(\theta - vy)(\Delta y)^2\varphi_x^3}{6v}A_{yyy} - \frac{\left[ vy\varphi_{xx} + 2\left(\frac{vy}{2}-r\right)\varphi_x^2\right](\Delta x)^2}{12} A_{xxx}\\
\\
&&  + \frac{\rho vy[(\Delta x)^2\varphi_x^2 -(\Delta y)^2]}{12}u_{xxxy} + \varepsilon .
\end{array}
\end{eqnarray}
For the \textit{Version 4 scheme} equations \eqref{uxxx}, \eqref{uyyy}, \eqref{uxxxx} and \eqref{uyyyy} are used in equation \eqref{semidiscretepdezudiskretisierenII} and then \eqref{uxxxy} is applied, which gives
\begin{eqnarray}\label{Version4}
\begin{array}{rcl}
\varphi_x^3 f &=& A_0 + \frac{vy(\Delta x)^2\varphi_x}{24} A_{xxxx} +\frac{vy(\Delta y)^2\varphi_x^3}{24}A_{yyyy}  + \frac{\rho vy(\Delta x)^2\varphi_x^2 }{12} A_{xxxy}\\
\\
&&+\frac{\kappa(\theta - vy)(\Delta y)^2\varphi_x^3}{6v}A_{yyy} - \frac{\left[ vy\varphi_{xx} + 2\left(\frac{vy}{2}-r\right)\varphi_x^2\right](\Delta x)^2}{12} A_{xxx}\\
\\
&&+ \frac{\rho vy\varphi_x^2 [(\Delta y)^2 - (\Delta x)^2\varphi_x^2]}{12}u_{xyyy} + \varepsilon .
\end{array}
\end{eqnarray}
\begin{remark}\label{remark1}
Equations \eqref{Version1}--\eqref{Version4} show that we can achieve a \textit{HOC} scheme
when either $\rho=0$, $v=0,$ or $(\Delta y)^2 \equiv (\Delta
x)^2\varphi_x^2$. 
The constraint $(\Delta y)^2 \equiv (\Delta x)^2\varphi_x^2$, however,
implies that the function $\varphi$ is affine linear and would not
qualify as a zoom function. In particular, the choice $\varphi(x)=x$
would yield the scheme discussed in \cite{DuFo12} (on a uniform grid), hence we will focus
on a zoom which is not affine linear.
\end{remark}

In equations \eqref{Version1} to \eqref{Version4} we observe that all
these schemes have a formal general consistency error of order two. But on the
other hand each version only has one remaining second order term, which is multiplied
with either $u_{xxxx}$, $u_{yyyy}$,
$u_{xxxy},$ or $u_{xyyy}$. All other terms are discretised with fourth
order accuracy. 
We call this an \textit{essentially high-order compact}
discretisation. To gauge the overall potential of the four discrete
schemes we obtain by neglecting the remaining second-order terms, it
is pivotal to understand the behaviour of these terms better.
To this end we compute a numerical solution using the (second-order)
central difference operator in $x$- and $y$-direction directly in
equation \eqref{pdezudiskretisierenII}, and obtain by numerical
differentiation (approximations of) the higher derivatives $u_{xxxx}$,
$u_{yyyy}$, $u_{xxxy}$, and $u_{xyyy}$ appearing in the remaining second order terms. 
\begin{center}
    \begin{minipage}[t]{.98\linewidth}
 \includegraphics[width=14cm,height=7.5cm]{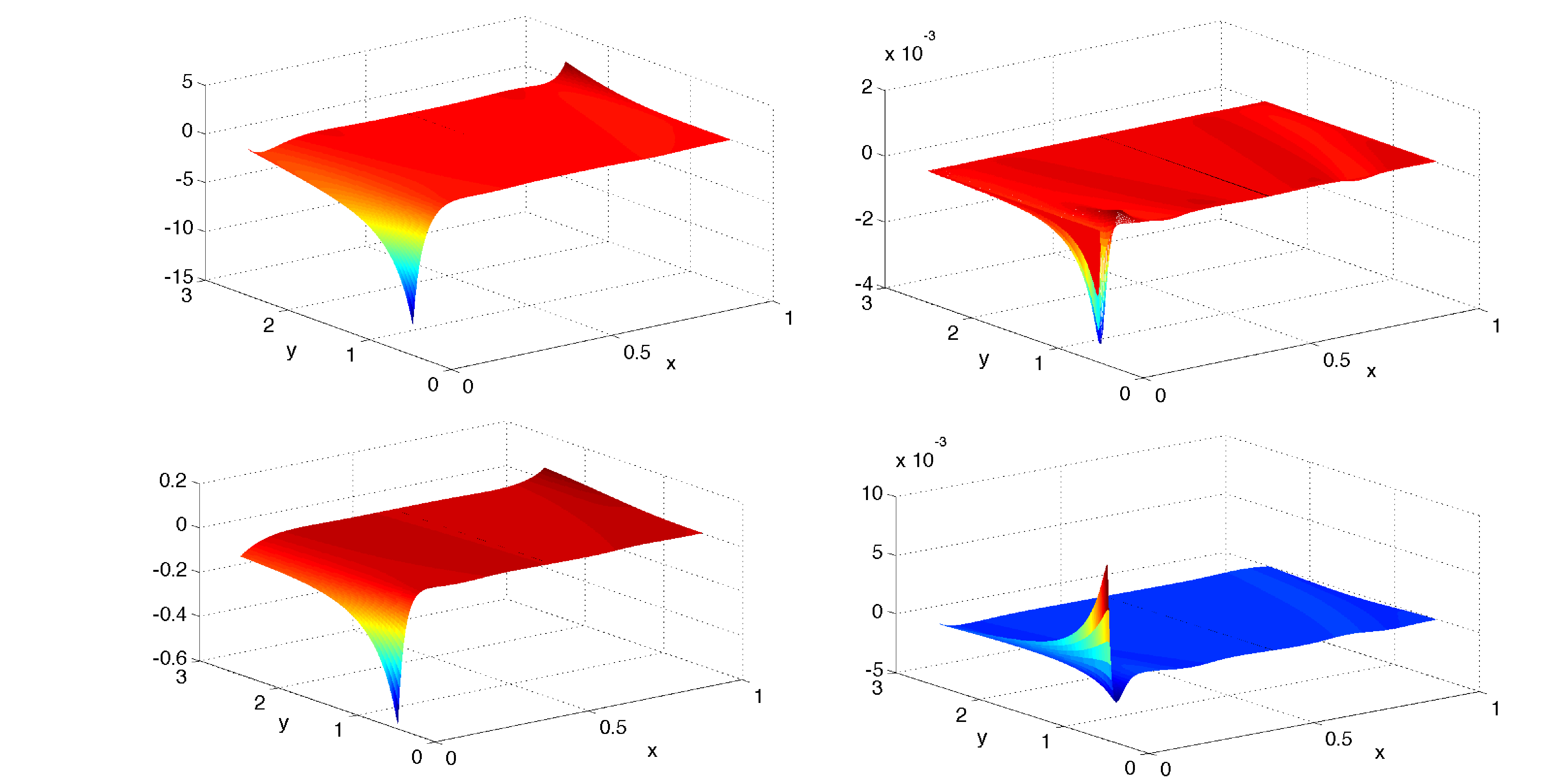}
	  \captionof{figure}{Remainder terms without $\mathcal{O}((\Delta x)^2)$ factor
            for Version 1 (upper left), Version 2 (upper right), Version 3 (lower left), and Version 4 (lower right)}
\label{fig:Remaining_Terms_All_Four_Schemes}
    \end{minipage}%
\end{center} 
Figure~\ref{fig:Remaining_Terms_All_Four_Schemes} shows the remainder
terms of second order appearing in equations
\eqref{Version1}--\eqref{Version4} without the $\mathcal{O}((\Delta
x)^2)$ factor, where $\rho =-0.1$, $\zeta = 2.5$, $p=1$, and
$S_{\min}=49.6694$. The values of these remainder terms determine if
we can achieve a fourth-order consistency, at least until a given
minimal step size. Hence, low values for the remainder terms are
favourable. We observe that all plots have in common that the highest
values of the remainder terms occur near the boundary $x=0$. 
On the upper left plot in
Figure~\ref{fig:Remaining_Terms_All_Four_Schemes} we see the remainder
term for Version 1. This term has by far the highest absolute
values. The $l^2$-norm of this remainder term is $8.8 \times
10^{-1}$. This indicates that a numerical study of this scheme
may not lead to a fourth-order consistency error. On the upper right
plot we have the remainder term for Version 2, again
without the $\mathcal{O}((\Delta
x)^2)$ factor. The highest absolute value for this is only
about $4\times 10^{-3}$, so very low when comparing it with the
remainder term of Version 1. The $l^2$-norm for this plot is
$3.1\times 10^{-4}$, which shows that Version 2 has a significantly
higher chance of producing a fourth order consistency error in the
numerical study than Version 1. The plot on the lower left side is
showing the remainder term of Version 3. This plot has higher values
than Version 2, but lower values than Version 1. With a $l^2$-norm of
$6.6\times 10^{-3}$ it has still a chance to produce a good
consistency error. The plot on the lower right shows the
remainder term of Version 4. This plot has again very low absolute
values which are only up to about $5\times 10^{-3}$. The $l^2$-norm
for this remainder term is $3.1 \times 10^{-4}$. This indicates that
we have a good chance that Version 4 produces a scheme with
fourth-order accuracy. 

In the special case that $\varphi(x)=x$ and $\Delta x = \Delta y=h$ we
have $(\Delta y)^2 \equiv (\Delta x)^2\varphi_x^2$, and all four
versions lead to exactly the same \textit{HOC\/} scheme,
\begin{equation*}
f = A_0 + \frac{vyh^2}{24} A_{xxxx} +\frac{vyh^2}{24}A_{yyyy}  + \frac{\rho vyh^2 }{12} A_{xxxy}+\frac{\kappa(\theta - vy)h^2}{6v}A_{yyy} - \frac{\left(\frac{vy}{2}-r\right)h^2}{6} A_{xxx} + \varepsilon ,
\end{equation*}
as in this case $A_{xxxy}=A_{xyyy}$ holds. This specific HOC scheme
without zoom is discussed in \cite{DuFo12}.

\begin{remark}\label{remark}
The derivation of the schemes in this section can be modified to
accommodate other stochastic 
volatility models as, e.g.\ the GARCH diffusion model
\eqref{eq:garchmodel} or the 3/2-model \eqref{eq:32model}.
Using these models the structure of the partial differential equation
\eqref{P0} remains the same, only the coefficients of
the derivatives have to be modified 
accordingly. Similarly, the coefficients of the derivatives in
\eqref{uxxx}-\eqref{uxyyy} have to be modified. Substituting
these in the modified expression for the truncation error one obtains equivalent
approximations as above.
\end{remark}

Our conclusion from the results in
Figure~\ref{fig:Remaining_Terms_All_Four_Schemes} is that Version 2
and Version 4 seem to be the best choices to obtain small
errors. The remainder term for Version 3 still has low values, while
Version 1 seems only to be able to produce a second-order scheme. 
Numerical experiments which we have carried out with all four
versions of the scheme indicate that actually Version 3 is leading to the best
results in terms of accuracy and stability.
Hence, in the remainder of this paper we focus on this particular
scheme. 

\section{High-order compact schemes for the parabolic problem}\label{sec:HOC_schemes_for_parabolic_problems}
We now consider the parabolic equation \eqref{pdezudiskretisierenII} with $f=-u_{\tau}$ and we denote by $U_{i,j}(\tau)$ the semi-discrete approximation of its solution $u(x_i,y_j,\tau)$ at time $\tau$.

\subsection{Semi-discrete schemes}
In this section we define the semi-discrete scheme of the form
\begin{equation}\label{semi_discrete_pde_two_dimensions_general}
\sum\limits_{\hat{z}\in G} \left[M_z(\hat{z}) \partial_{\tau} U_{i,j}(\tau) + K_z(\hat{z}) U_{i,j}(\tau)  \right] =  0,
\end{equation}
at time $\tau$ for each point $z\in  \interior{G}$, where
$\interior{G}$ denotes the inner points of the grid $G$. We use
$\Delta x = \Delta y = h $ for some $h>0$ in the definition of
$G$, which is given in \eqref{Definition_of_general_grid}. We have that
$K_z(\hat{z})$ and $M_z(\hat{z})$ are operators with nine values
defined on the compact stencil around $z\in \interior{G}$. 

Using the central difference operator in \eqref{Version3} at the point $z\in \interior{G}$ leads to
\begin{eqnarray}\label{Scheme_Heston_Model_Version_3_K_equation_one}
\begin{array}{rcl}
\hat{K}_{i+1,j\pm 1} & = & 
\frac {{\varphi^{4} _{{x}}} \left( \frac{vy}{2}-r \right) }{24h} 
- \frac {vy{\varphi^{2} _{{x}}}\varphi _{{{\it xx}}}}{16h} 
+ \frac { \left( \frac{vy}{2}-r \right) {\varphi^{2} _{{x}}}}{24h} 
+ \frac {vy\varphi _{{{\it xx}}}}{48h} 
- \frac {vy\varphi _{{x}}}{24{h}^{2}} 
- \frac {vy{\varphi^{3} _{{x}}}}{24{h}^{2}} 
\mp \frac { \varphi _{{x}}\kappa\, \left( \theta-vy \right) }{24vh}\\
\\
&& 
\mp \frac {\kappa\,{\varphi^{3} _{{x}}} \left( \theta-vy \right) }{24vh} 
\pm 
\frac {\kappa\, \left( \theta-vy \right)  \left( \frac{vy}{2}-r \right) {\varphi^{2} _{{x}}}}{24{v}^{2}y} 
\pm \frac {\kappa\, \left( \theta-vy \right) \varphi _{{{\it xx}}}}{48v} 
\mp \frac { \left( \frac{vy}{2}-r \right) {\varphi^{2} _{{x}}}}{24y} 
 \pm \frac{v{\varphi^{2} _{{x}}}}{48} \\
 \\
 && 
\pm \frac {{\varphi^{4} _{{x}}}\kappa\, \left( \theta-vy \right)  \left( \frac{vy}{2}-r \right) }{24{v}^{2}y} 
\mp \frac {\kappa\, \left( \theta-vy \right) {\varphi^{2} _{{x}}}\varphi _{{{\it xx}}}}{16v} 
+ \rho^2 \left[ 
\frac {vy\varphi _{{{\it xx}}}}{12h}  
\pm \frac{v\varphi _{{{\it xx}}}}{8} 
- \frac {vy\varphi _{{x}}}{6{h}^{2}} 
\right] \\
 \\
 && 
+ \rho \left[
\pm \frac{{\varphi^{2} _{{x}}}\varphi _{{{\it xx}}} \left( \frac{vy}{2}-r \right)}{12} 
\pm \frac{vy{\varphi^{2} _{{{\it xx}}}}}{12} 
\mp \frac{ \left( \frac{vy}{2}-r \right) \varphi _{{{\it xx}}}}{24} 
\pm \frac {vy\varphi _{{{\it xxx}}}}{48\varphi _{{x}}} 
\pm \frac {\varphi _{{x}} \left( \frac{vy}{2}-r \right) }{12h}  
\pm \frac {{\varphi^{3} _{{x}}} \left( \frac{vy}{2}-r \right) }{12h} 
\right.\\
 \\
 && \left. 
\pm \frac {vy\varphi _{{{\it xx}}}}{8h\varphi _{{x}}} 
\pm \frac{{\varphi^{2} _{{x}}}\kappa}{24}  
\mp \frac {vy\varphi _{{x}}\varphi _{{{\it xx}}}}{24h} 
- \frac {{\varphi^{2} _{{x}}}\kappa\, \left( \theta-vy \right) }{6hv} 
\mp \frac {vy{\varphi^{2} _{{{\it xx}}}}}{16{\varphi^{2} _{{x}}}} 
\mp \frac {vy{\varphi^{2} _{{x}}}}{4{h}^{2}} 
\pm \frac {v{\varphi^{2} _{{x}}}}{24y}  \mp \frac{vy\varphi _{{x}}\varphi _{{{\it xxx}}}}{24} 
\right] ,
\end{array}
\end{eqnarray}
\begin{eqnarray}\label{Scheme_Heston_Model_Version_3_K_equation_two}
\begin{array}{rcl}
\hat{K}_{i-1,j\pm 1} & = &- \hat{K}_{i+1,j\pm 1} 
- \frac {vy\varphi _{{x}}}{12{h}^{2}} 
- \frac {vy{\varphi^{3} _{{x}}}}{12{h}^{2}} 
\mp \frac {\varphi _{{x}}\kappa\, \left( \theta-vy \right) }{12vh} 
\mp \frac {{\varphi^{3} _{{x}}} \kappa\, \left( \theta-vy \right) }{12vh} 
- \rho^2 \frac { vy\varphi _{{x}}}{3{h}^{2}} 
\\
\\
&&
+\rho \left[
\pm \frac {\varphi _{{x}} \left( \frac{vy}{2}-r \right) }{6h}
\pm \frac {vy\varphi _{{{\it xx}}}}{4h\varphi _{{x}}}  
 \pm \frac {{\varphi^{3} _{{x}}} \left( \frac{vy}{2}-r \right) }{6h} 
 \mp \frac {vy\varphi _{{x}}\varphi _{{{\it xx}}}}{12h}  
\right],
\end{array}
\end{eqnarray}
\begin{eqnarray}\label{Scheme_Heston_Model_Version_3_K_equation_three}
\begin{array}{rcl}
\hat{K}_{i\pm 1,j} & = & 
\frac {vy{\varphi^{3} _{{x}}}}{12{h}^{2}} 
\mp \frac{h{\varphi^{2} _{{{\it xx}}}} \left( \frac{vy}{2}-r \right)}{6}  
\mp \frac {{\varphi^{4} _{{x}}} \left( \frac{vy}{2}-r \right) }{12h} 
\pm \frac { 5\left( \frac{vy}{2}-r \right) {\varphi^{2} _{{x}}}}{12h} 
\pm \frac{yhv\varphi _{{{\it xxxx}}}}{48} 
\mp \frac {h\varphi _{{{\it xx}}}v}{24y} \\
\\
&& 
- \frac {\varphi _{{x}}\kappa\, \left( \theta-vy \right) }{12vy} 
- \frac {5vy\varphi _{{x}}}{12{h}^{2}} 
\pm \frac {5vy\varphi _{{{\it xx}}}}{24h} 
+ \frac {v\varphi _{{x}}}{12y} 
\mp \frac {{\varphi^{2} _{{x}}}hv}{24y} 
- \frac {{\varphi^{3} _{{x}}} \left( \frac{vy}{2}-r \right) ^{2}}{6vy} 
+ \frac{vy\varphi _{{{\it xxx}}}}{24} \\
\\
&&
\pm \frac{\varphi _{{x}}h \left( \frac{vy}{2}-r \right) \varphi _{{{\it xxx}}}}{24} 
\pm \frac {vy{\varphi^{2} _{{x}}}\varphi _{{{\it xx}}}}{8h}  
+ \frac{ \left( \frac{vy}{2}-r \right) \varphi _{{x}}\varphi _{{{\it xx}}}}{12} 
\mp \frac {vyh\varphi _{{{\it xx}}}\varphi _{{{\it xxx}}}}{16\varphi _{{x}}} 
\pm \frac {h\kappa\, \left( \theta-vy \right) \varphi _{{{\it xx}}}}{24vy} 
\\
\\
&& 
\mp \frac {{\varphi^{2} _{{x}}}h \left( \frac{vy}{2}-r \right) ^{2}\varphi _{{{\it xx}}}}{6vy} 
\pm \frac {{\varphi^{2} _{{x}}}h\kappa\, \left( \theta-vy \right) }{24vy} 
+ \rho^2 \left[ 
\frac {vy\varphi _{{x}}}{3{h}^{2}} 
\mp \frac {vy\varphi _{{{\it xx}}}}{6h} 
\right] \\
\\
&& 
+ \rho \left[ 
\frac {v\varphi _{{{\it xx}}}}{4\varphi _{{x}}} 
\mp \frac{h\varphi _{{{\it xx}}}v }{24} 
\mp \frac {hv{\varphi^{2} _{{{\it xx}}}}}{8{\varphi _{{x}}}^{2}} 
+ \frac{v\varphi _{{x}}}{12} 
- \frac {\varphi _{{x}} \left( \frac{vy}{2}-r \right) }{6y} 
\mp \frac {h \left( \frac{vy}{2}-r \right) \varphi _{{{\it xx}}}}{6y} 
\pm \frac {{\varphi^{2} _{{x}}}\kappa\, \left( \theta-vy \right) }{3hv} 
\right],
\end{array}
\end{eqnarray}
\begin{eqnarray}\label{Scheme_Heston_Model_Version_3_K_equation_four}
\begin{array}{rcl}
\hat{K}_{i,j\pm 1} & = &  
\frac{{\varphi^{3} _{{x}}}\varphi _{{{\it xx}}} \left( \frac{vy}{2}-r \right)}{4} 
\pm \frac {{\varphi^{3} _{{x}}}h \left( \frac{vy}{2}-r \right) \kappa\, \left( \theta-vy \right) \varphi _{{{\it xx}}}}{4{v}^{2}y} 
\mp \frac {{\varphi^{2} _{{x}}}h\kappa\, \left( \theta-vy \right) \varphi _{{{\it xxx}}}}{8v} 
 - \frac {5vy{\varphi^{3} _{{x}}}}{12{h}^{2}} 
+ \frac {{\varphi^{3} _{{x}}}v}{12y} 
\\
\\
&& 
- \frac {{\varphi^{3} _{{x}}}{\kappa}^{2} \left( \theta-vy \right) ^{2}}{6y{v}^{3}} 
+ \frac {vy\varphi _{{x}}}{12{h}^{2}} 
\mp \frac {{\varphi^{3} _{{x}}}h\kappa}{12y} 
\pm \frac {{\varphi^{3} _{{x}}}h{\kappa}^{2} \left( \theta-vy \right) }{12{v}^{2}y} 
\mp \frac {5\kappa\,{\varphi^{3} _{{x}}} \left( \theta-vy \right) }{12vh} 
+ \frac{vy\varphi _{{x}}{\varphi^{2} _{{{\it xx}}}}}{8} 
\\
\\
&& 
+ \frac {\kappa\,{\varphi^{3} _{{x}}} \left( \theta-vy \right) }{12vy} 
+ \frac{\kappa\,{\varphi^{3} _{{x}}}}{6} 
\pm \frac {\varphi _{{x}}\kappa\, \left( \theta -vy \right) }{12vh} 
\pm \frac {\varphi _{{x}}h{\varphi^{2} _{{{\it xx}}}}\kappa\, \left( \theta-vy \right) }{8v} 
- \frac{vy{\varphi^{2} _{{x}}}\varphi _{{{\it xxx}}}}{8} 
+ \rho^2 \frac {vy\varphi _{{x}}}{3{h}^{2}} 
\\
\\
&& 
+ \rho \left[ 
\pm \frac {vy\varphi _{{x}}\varphi _{{{\it xx}}}}{12h} 
\mp \frac {{\varphi^{3} _{{x}}} \left( \frac{vy}{2}-r \right) }{6h} 
\pm \frac {h
\varphi _{{x}}\kappa\, \left( \theta-vy \right) \varphi _{{{\it xx}}}}{4vy} 
\mp \frac {vy\varphi _{{{\it xx}}}}{4h\varphi _{{x}}} 
+ \frac{v\varphi _{{x}}\varphi _{{{\it xx}}}}{4} 
\mp \frac {\varphi _{{x}} \left( \frac{vy}{2}-r \right) }{6h} 
\right]
\end{array}
\end{eqnarray}
and
\begin{eqnarray}\label{Scheme_Heston_Model_Version_3_K_equation_five}
\begin{array}{rcl}
\hat{K}_{i,j} & = & 
\frac{vy{\varphi^{2} _{{x}}}\varphi _{{{\it xxx}}}}{4} 
- \frac{{\varphi^{3} _{{x}}}\varphi _{{{\it xx}}} \left( \frac{vy}{2}-r \right)}{2} 
- \frac{vy\varphi _{{x}}{\varphi^{2} _{{{\it xx}}}}}{4}  
- \frac {{\varphi^{3} _{{x}}}v}{6y} 
- \frac {{\varphi^{3} _{{x}}} \kappa\, \left( \theta-vy \right) }{6vy} 
- \frac{\kappa\,{\varphi^{3} _{{x}}}}{3} \\
\\
&& 
+ \frac {{\varphi^{3} _{{x}}}{\kappa}^{2} \left( \theta-vy \right) ^{2}}{3y{v}^{3}} 
+ \frac {5vy\varphi _{{x}}}{6{h}^{2}} 
+ \frac {5vy{\varphi^{3} _{{x}}}}{6{h}^{2}} 
- \frac{ \left( \frac{vy}{2}-r \right) \varphi _{{x}}\varphi _{{{\it xx}}}}{6} 
- \frac{vy\varphi _{{{\it xxx}}}}{12} 
+ \frac {{\varphi^{3} _{{x}}} \left( \frac{vy}{2}-r \right) ^{2}}{3vy} 
\\
\\
&& 
- \frac {v\varphi _{{x}}}{6y} 
+ \frac {\varphi _{{x}}\kappa\, \left( \theta-vy \right) }{6vy} 
- \rho^2 \frac {2vy\varphi _{{x}}}{3{h}^{2}}  
+ \rho \left[ 
\frac {\varphi _{{x}} \left( \frac{vy}{2}-r \right) }{3y} 
- \frac {v\varphi _{{{\it xx}}}}{2 \varphi _{{x}}} 
- \frac{v\varphi _{{x}}}{6} 
- \frac{v\varphi _{{x}}\varphi _{{{\it xx}}}}{2} 
\right] ,
\end{array}
\end{eqnarray}
where $\hat{K}_{i,j}$ is the coefficient of $U_{i,j}(\tau)$. For the sake of readability we drop the subindex $i$ on the derivatives of $\varphi$ and the subindex $j$ on $y$, respectively. Analogously we have
\begin{eqnarray}\label{Scheme_Heston_Model_Version_3_M1_to_M9}
\begin{array}{rcl}
\hat{M}_{i+1,j\pm 1} & = &\hat{M}_{i-1,j \mp 1} = 
\pm \rho \frac{{\varphi^{2} _{{x}}}}{24},\\
\\
\hat{M}_{i,j\pm 1} & = &   
\frac{{\varphi^{3} _{{x}}}}{12} 
\mp \frac {{\varphi^{3} _{{x}}}h}{12y} 
\pm \frac {{\varphi^{3} _{{x}}}h\kappa\, \left( \theta-vy \right) }{12{v}^{2}y} ,\\
\\
\hat{M}_{i\pm 1,j} & = &  
\frac{{\varphi^{3} _{{x}}}}{12} 
\mp \frac {{\varphi^{4} _{{x}}}h \left( \frac{vy}{2}-r \right) }{12vy} 
\pm \frac{{\varphi^{2} _{{x}}}h\varphi _{{{\it xx}}}}{8}
\mp \rho \frac {{\varphi^{2} _{{x}}}h}{12y} \text{ and }\\
\\
\hat{M}_{i,j} & = & 
\frac{2{\varphi^{3} _{{x}}}}{3} 
- \frac {{\varphi^{3} _{{x}}}{h}^{2}\varphi _{{{\it xx}}} \left( \frac{vy}{2}-r \right) }{2vy} 
- \frac{\varphi _{{x}}{h}^{2}{\varphi^{2} _{{{\it xx}}}}}{4} 
+ \frac{{\varphi^{2} _{{x}}}\varphi _{{{\it xxx}}}{h}^{2}}{4}  
- \rho \frac {\varphi _{{x}}\varphi _{{{\it xx}}}{h}^{2}}{2y},
\end{array}
\end{eqnarray}
as coefficients of $\partial_{\tau}U_{i,j}(\tau) $. With the usage of $z \in \interior{G}$ we have 
\begin{eqnarray}\label{equation_defining_eq_system_with_coefficients_K_hat_and_M_hat}
 \begin{array}{rcl}
K_{z}(\hat{z})  =\hat{ K}_{n_1,n_2} &\text{ as well as }& M_{z}(\hat{z})  =\hat{ M}_{n_1,n_2} 
\end{array}
\end{eqnarray}
for $$\hat{z} = \left(x_{n_1},y_{n_2} \right)$$ with $n_1 \in \{ i -1, i, i+1\}$ and  $n_2 \in \{ j -1, j, j+1\}$.
Thus \eqref{semi_discrete_pde_two_dimensions_general} corresponds to a linear system on $\interior{G}$.

\subsection{Treatment of the boundary conditions}\label{sec:boundary_conditions}
The first boundary is the \textit{boundary }${x=x_{\min}}$,
which corresponds to the boundary at $S=0$ of the
original problem. For this boundary 
we have to discount the option price at time $T$ to the appropriate
time. Taking into account the transformations $\tau=T-t$ and
$u=e^{r\tau}{V}/{K}$ this leads to the Dirichlet
boundary condition 
\begin{eqnarray}
\notag u(x_{\min},y,\tau)=u(x_{\min},y,0) \text{ for all } \tau \in [0,\tau_{\max}] \text{ and all } y \in [y_{\min},y_{\max}]. 
\end{eqnarray}

The next boundary we discuss is the \textit{boundary
}${x=x_{\max}}$, which corresponds to the boundary at
$S=S_{\max}$ of the original problem.
For a Power Put with power $p$ we have
\begin{eqnarray}
\notag \lim\limits_{S \rightarrow \infty} V(S,\sigma,t)=0,
\end{eqnarray}
which we approximate at the artificial boundary $S_{\max}$ by $V_{S}(S_{\max}, \sigma, t)=0$, $V_{SS}(S_{\max}, \sigma, t)=0$, $V_{S\sigma}(S_{\max}, \sigma, t)=0$, $V_{\sigma}(S_{\max}, \sigma, t)=0$ as well as $V_{\sigma\sigma}(S_{\max}, \sigma, t)=0$. Using these approximations in \eqref{Heston_model_original_pde} gives
\begin{eqnarray}
\notag V_t  -rV=0.
\end{eqnarray}
Using $\tau=T-t$ and $u=e^{r\tau}{V}/{K}$ yields $u_{\tau} =0$ and thus the Dirichlet boundary condition
\begin{eqnarray}
u(x_{\max},y, \tau)= u(x_{\max},y, 0)  \text{ for all } \tau \in [0,\tau_{\max}]\text{ and all } y \in [y_{\min},y_{\max}]. 
\end{eqnarray}

The third boundary to discuss is the \textit{boundary
}${y=y_{\min}}$ with $x \notin \{x_{\min},x_{\max}\}$, which
corresponds to the boundary $\sigma=\sigma_{\min}$ with $S \notin
\{S_{\min},S_{\max}\}$. We will treat
this boundary just like the inner of the computational domain, using
the equations \eqref{Scheme_Heston_Model_Version_3_K_equation_one}
to \eqref{Scheme_Heston_Model_Version_3_K_equation_five}. This
requires the usage of ghost-points $U_{i-1,-1}$, $U_{i,-1}$ and
$U_{i+1,-1}$ when discretising at the points $(x_i,y_0)\in G$ for
$i=1, \ldots, N-1$. So we need a fourth order accurate expression for
the ghost-points $U_{i,-1}$ for $i=0, \ldots, N$. We use the following
extrapolation formula
\begin{eqnarray}
\notag U_{i,-1}= 4 U_{i,0} - 6 U_{i,1} + 4U_{i,2} - U_{i,3} + \mathcal{O}\left((\Delta y)^4\right)
\end{eqnarray}
for $i=0, \ldots , N$. The same procedure is used for the
ghost-points for the matrix $M_h$ when using the equations in
\eqref{Scheme_Heston_Model_Version_3_M1_to_M9}.

The last boundary we discuss is the boundary at \textit{boundary
}${y=y_{\max}}$ with $x \notin \{x_{\min},x_{\max}\}$, which is
corresponding to the boundary $\sigma=\sigma_{\max}$ with $S \notin
\{S_{\min},S_{\max}\}$ of the untransformed problem. We treat this
boundary similar as the boundary at $y_{\min}$ and use equations
\eqref{Scheme_Heston_Model_Version_3_K_equation_one} to
\eqref{Scheme_Heston_Model_Version_3_K_equation_five}. The scheme
then uses, when discretising at the points $(x_i,y_M)\in G$ for $i=1, \ldots, N-1$, the ghost-points $U_{i-1,M+1}$, $U_{i,M+1}$ and $U_{i+1,M+1}$
for $i=1, \ldots, N-1$. This means that we have to find an expression for the ghost-points $U_{i,M+1}$, $i=0, \ldots, N$. We approximate the values at these
ghost-points again using extrapolation, 
\begin{eqnarray}
\notag U_{i,M+1} = 4 U_{i,M} - 6 U_{i,M-1} + 4 U_{i,M-2} -  U_{i,M-3} + \mathcal{O}\left((\Delta y)^4\right)
\end{eqnarray}
for $i=0, \ldots , N$. Again, the same procedure is used for the
ghost-points for the matrix $M_h$ while using the equations in
\eqref{Scheme_Heston_Model_Version_3_M1_to_M9}. 

\subsection{Time discretization}
With the results from the previous sections we obtain a semi-discrete system of the form
\begin{equation}\label{semi_discrete_pde_two_dimensions_general_including_boundaries}
\begin{array}{rcl}
\sum\limits_{\hat{z}\in G} \left[M_z(\hat{z}) \partial_{\tau} U_{i,j}(\tau) + K_z(\hat{z}) U_{i,j}(\tau) \right]& = & g(z)
\end{array}
\end{equation}
for each point $z$ of the grid $G$, which is defined in \eqref{Definition_of_general_grid} and $\Delta x = \Delta y = h $ for some $h>0$ is used. The function $g(z)$ has only non-zero values at the boundaries $x_{\min}$ and $x_{\max}$.

We use a time grid of the form $$ \left\{ \frac{\Delta \tau}{4}
  ,\frac{\Delta \tau}{2} ,\frac{3\Delta \tau}{4} ,\Delta \tau ,2\Delta
  \tau,3\Delta \tau, \ldots \right\},$$where the first time steps have
step size $\frac{\Delta \tau}{4}$ and the following have $\Delta
\tau$. For these first four time steps, we use the implicit Euler
scheme, and obtain
\begin{eqnarray}
\notag \begin{array}{rcl} 
\sum\limits_{\hat{z}\in G} \left[ M_z(\hat{z}) + \frac{\Delta \tau}{4}K_z(\hat{z})\right] U_{i,j}^{n+1}  &=& \sum\limits_{\hat{z}\in G}  M_z(\hat{z})U_{i,j}^{n}  + \frac{\Delta \tau}{4}g(z)
\end{array}
\end{eqnarray}
with $n=0,1,2,3$ for each grid-point $z \in G$. This approach is suggested in \cite{Ran84} when dealing with
non-smooth initial conditions. For the following time steps we use a
Crank-Nicolson-type time discretisation, leading to
\begin{eqnarray}
\notag \begin{array}{rcl}
\sum\limits_{\hat{z}\in G} \left[ M_z(\hat{z}) + \frac{\Delta \tau}{2}K_z(\hat{z})\right] U_{i,j}^{n+1}  &=& \sum\limits_{\hat{z}\in G} \left[ M_z(\hat{z}) - \frac{\Delta \tau}{2}K_z(\hat{z})\right]U_{i,j}^{n+1}  + (\Delta \tau)g(z)
\end{array}
\end{eqnarray}
with $n \geq 4$ on each point $z$ of the grid $G$. We observe that we
have only non-zero values on the compact computational stencil as $M_x(\hat{x})$ and $K_x(\hat{x})$ have this property. For the Crank-Nicolson time discretisation this compact scheme has consistency
order two in time and four in space for $\varphi(x)=x$ and $\rho=0$ or
is essentially high-order compact in space otherwise.
%
%

\section{Numerical Experiments}\label{Application_of_Version_3_Heston_with_zoom}
In this section we present the results of our numerical experiments for the compact
scheme using \eqref{Scheme_Heston_Model_Version_3_K_equation_one} -
\eqref{Scheme_Heston_Model_Version_3_M1_to_M9}, whose boundary conditions were derived in Section~\ref{sec:boundary_conditions}. 
If not stated otherwise, we will use the following default model parameters
\begin{equation*}
  \kappa = 1.1,\quad \theta = 0.15,\quad v = 0.1,\quad r=\ln(1.05),\quad K=100, \quad T=0.25.
\end{equation*}
The initial condition for the European (Power) Put after transformation
as in Section~\ref{sec:trans} is given by
\begin{eqnarray}\label{Initial_condition_transformed_problem}
u(x,y,0)= K^{p-1}\max \left(1-e^{\varphi(x)},0\right)^{p},
\end{eqnarray}
where the non-differentiable point of the initial condition is at
$x_K=\varphi^{-1} (0)$.

\subsection{Choice of the zoom function}

In our numerical experiments we use the zoom function
\begin{eqnarray}\label{zoomfunctionTANGMANN}
\hat{S}=\varphi(x)= \frac{\sinh(c_2x+c_1(1-x))}{\zeta},
\end{eqnarray}
proposed in \cite{TaGoBh08}, with $c_1= \asinh(\zeta \hat{S}_{\min})$,
$ c_2= \asinh(\zeta \hat{S}_{\max})$ and $\zeta>0$. The
non-differentiable point of the initial condition hence is at
\begin{eqnarray}
\notag x_K=\varphi^{-1} (0)=\frac{\asinh(0) - c_1}{c_2-c_1}=\frac{- \asinh(\zeta \hat{S}_{\min})}{\asinh(\zeta \hat{S}_{\max})-\asinh(\zeta \hat{S}_{\min})}.
\end{eqnarray}
Using the definitions of $c_1$ and $c_2$ this can be rearranged to
\begin{eqnarray}\label{x_position_des_Knicks_in_Initial_Condition_hatSmin_gleich_minus_hatSmax}
\hat{S}_{\min}=\frac{\sinh\left(\frac{x_K}{x_K-1}\asinh(\zeta \hat{S}_{\max}) \right)}{\zeta}.
\end{eqnarray}
Hence, $\hat{S}_{\min}$ can be set by choosing $x_K$ in reasonable
bounds as well as choosing $S_{\max}$, which gives $\hat{S}_{\max}$,
for a given $\zeta$. The fact that $x_K$ can be chosen is very
helpful, since if the non-differentiable point is on the grid the
numerical convergence order may be reduced to two in practice. 
Hence, we choose the grid such that the point $x_K$ in the middle of
two consecutive grid points on the finest grid. This
procedure of shifting the grid has
been suggested, for example, in \cite{Tavella}.

In the numerical experiments reported below we choose
\begin{equation*}
S_{\min} =Ke^{\hat{S}_{\min}}, \quad S_{\max} = 2K, \quad
\sigma_{\min} =0.05, \quad  \sigma_{\max}=0.25. 
\end{equation*}

\begin{figure}
  \centering
  \includegraphics[width=0.6\textwidth]{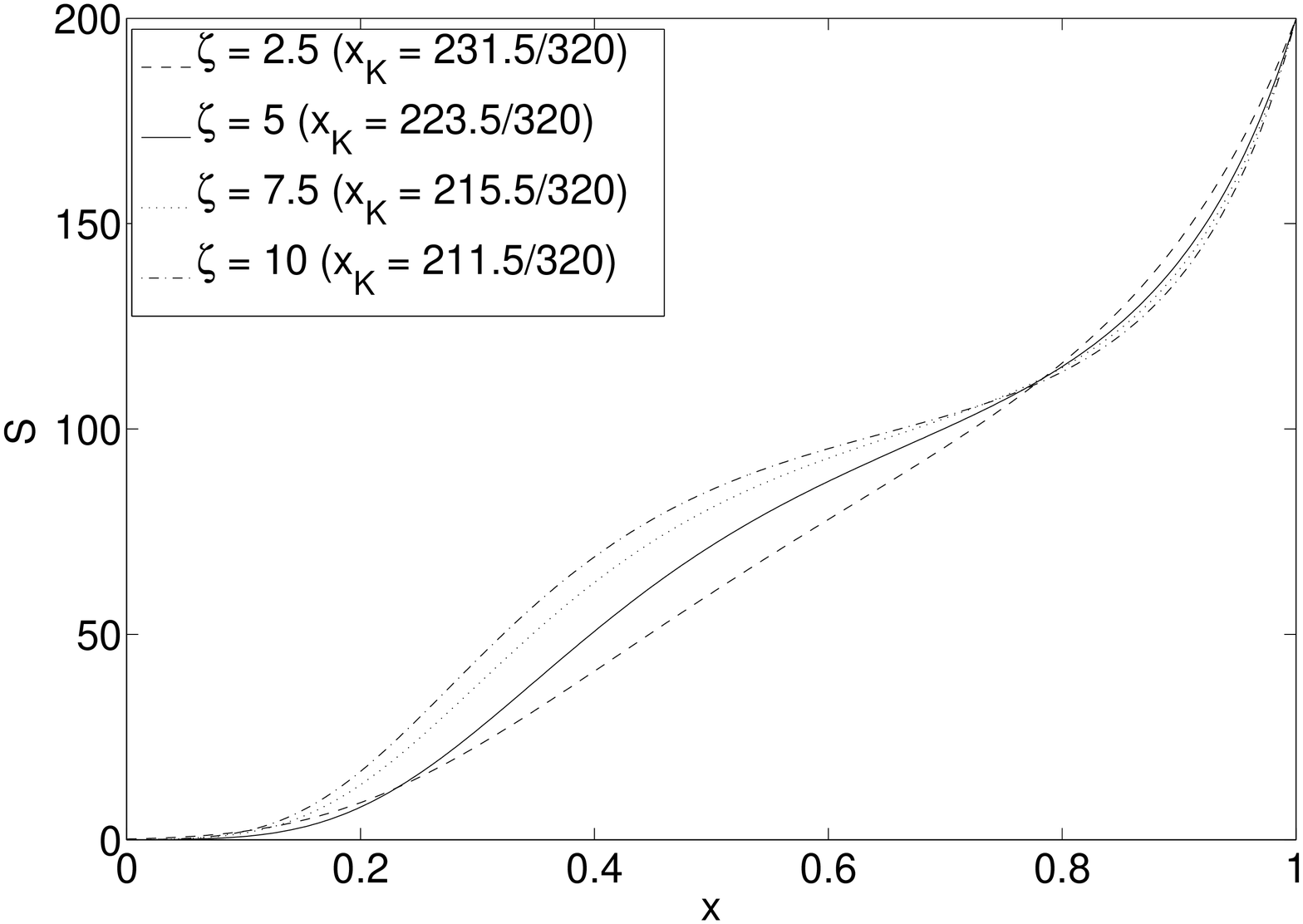}
  \caption{Different zoom examples with $K=100.$}
  \label{fig:zoomfunction_Tangman_examples_with_different_zeta}
\end{figure}

Figure~\ref{fig:zoomfunction_Tangman_examples_with_different_zeta}
shows the influence of the parameter $\zeta$ on the zoom in equation
\eqref{zoomfunctionTANGMANN},  taking into account both transformations, $\hat{S} = \ln \left({S}/{K}\right)$ and $x=\varphi^{-1} (\hat{S}).$
The different values for $x_K$, which depends on $\zeta$, are
chosen in such a way that the focus on the values
around $S=0$ is not too pronounced, compare equation
\eqref{x_position_des_Knicks_in_Initial_Condition_hatSmin_gleich_minus_hatSmax}. We
observe that for smaller values of $\zeta>0$ there is less zoom. So
with $\zeta \rightarrow 0$ the zoom function is approaching the
linear transformation $\varphi(x)=
(\hat{S}_{\max}-\hat{S}_{\min})x + \hat{S}_{\min}$ with $x
\in [0,1]$. With a larger value of $\zeta$ there is a stronger focus on our area of interest around the exercise
price $K$. 

The aim is to find an `optimal' value for $\zeta$ to be used in
practical computations.
The larger $\zeta$, the smaller the error around $K$, but on the other
hand the error in other parts of the domain increases when having a
stronger zoom, because an increasing number of grid points in the area
around $K$ automatically results into a decreasing amount of
grid points in other areas and vice versa. There has to be a balance between the
error in the area around
$K$ and the error in other parts of the domain. The overall order of
convergence should be looked at to achieve this balance and thus to
get a good value for $\zeta$. We expect the numerical convergence order to
increase at first with rising $\zeta$ and then decrease again after a
certain `optimal' strength of zoom is reached.

\subsection{Numerical convergence}

We now study the numerical errors of the discretisation as $h\to 0$ for fixed parabolic mesh ratio $\Delta
\tau /h^2,$ using different 
values for $\zeta$ and $\rho$. We compute an approximation of
the solution of the transformed problem, which is given by equation
\eqref{pdezudiskretisierenII}, and then transform it back into the
original variables. For
the relative $l^2$- and $l^{\infty}$-error plots a reference solution is
computed on a fine grid with $h_{\text{ref}} = 0.003125$. For the
relative $l^2$-errors we use 
$$\frac{\Vert U_{\text{ref}} - U\Vert_{l^2}}{\Vert U_{\text{ref}}\Vert_{l^2}}$$ 
and for the
$l^{\infty}$-error we use 
$$\Vert U_{\text{ref}} - U \Vert_{l^{\infty}},$$ 
where $U_{\text{ref}}$ denotes the reference solution and $U$ is the approximation. We
expect the error to behave like $\mathcal{O}\left(h^k \right)$ for
some $k$. If we plot the logarithm of the
error against the logarithm of the number of grid points,
the slope of this log-log plot gives the numerical convergence order of
the scheme.  Due to the initial condition of the transformed
problem not being smooth everywhere, we observe that the log-log plots
do not always produce a straight line, e.g.\ for a plain vanilla Put
option. For a smooth initial condition the log-log plots of the
errors give an almost straight line, e.g.\ for the Power
Put option. The numerical convergence order indicated in the figures
below is always computed as the slope of the linear least square fit
of the error points. For comparison we additionally plot the results
for a standard discretisation (SD), which means that the standard
central difference operator is used in \eqref{pdezudiskretisierenII}
as well as  
\begin{eqnarray}
\notag \varphi (x) = \left( \hat{S}_{\max} - \hat{S}_{\max}\right)x + \hat{S}_{\min}.
\end{eqnarray}
In this way all discretisations considered here operate on the same
spatial grid and a meaningful comparison can occur. We use
$\Delta \tau = 0.4 h^2$ for all convergence plots, although we note
that the dependence of the numerical convergence order on the
choice of the parabolic mesh ratio is marginal. This is in line with
the results of our numerical stability study reported below in Section~\ref{sec:stabstudy}.
\begin{center}
    \begin{minipage}[t]{.49\linewidth}
      \includegraphics[width=8cm,height=5.5cm]{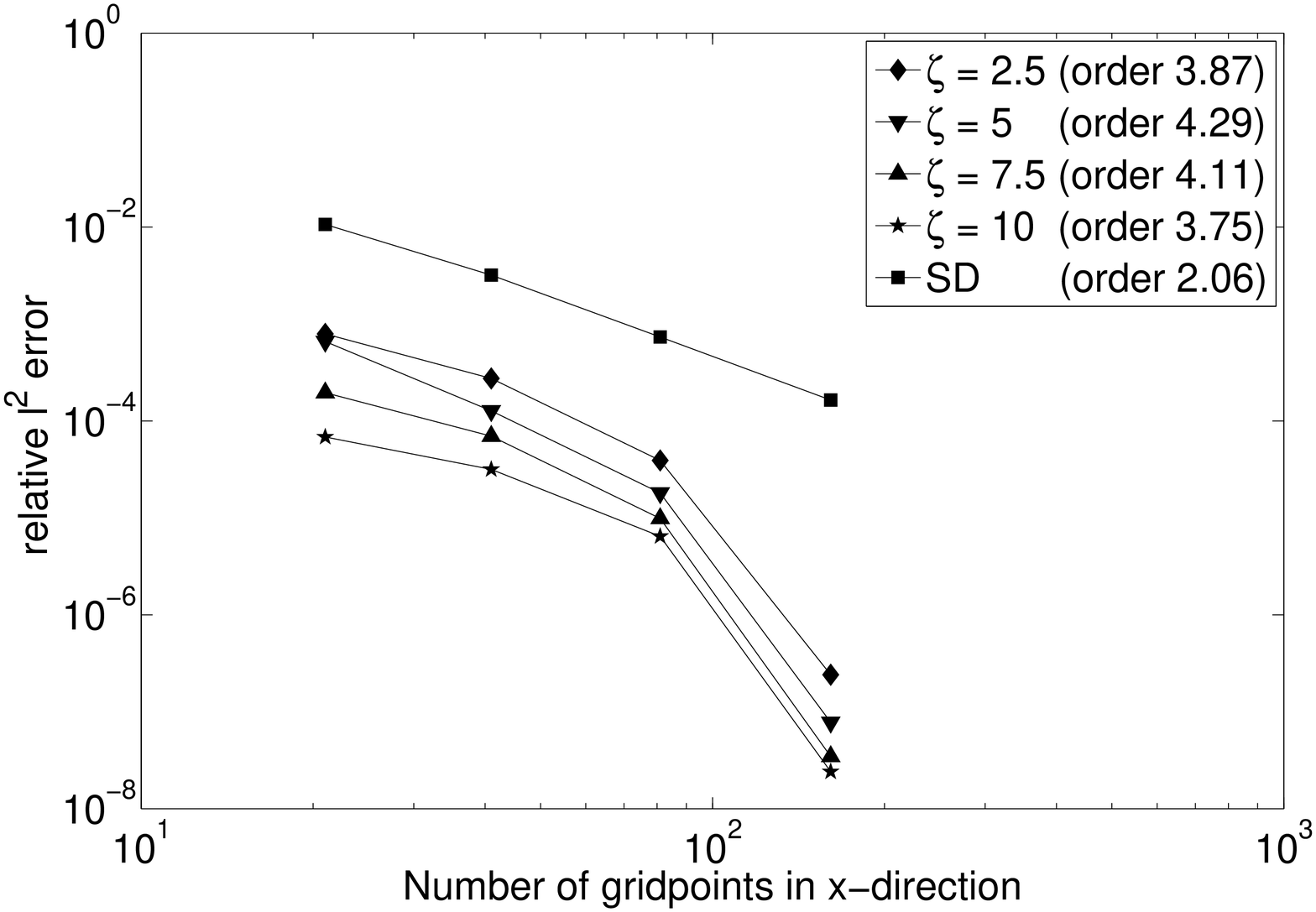}
	  \captionof{figure}{Relative $l^2$-error Heston model $\rho=0$}
\label{fig:loglog_plot_L_2_error_Heston_Hull_White}
    \end{minipage}%
    \begin{minipage}{.02\linewidth}
    ~
    \end{minipage}%
    \begin{minipage}[t]{.49\linewidth}
      \includegraphics[width=8cm,height=5.5cm]{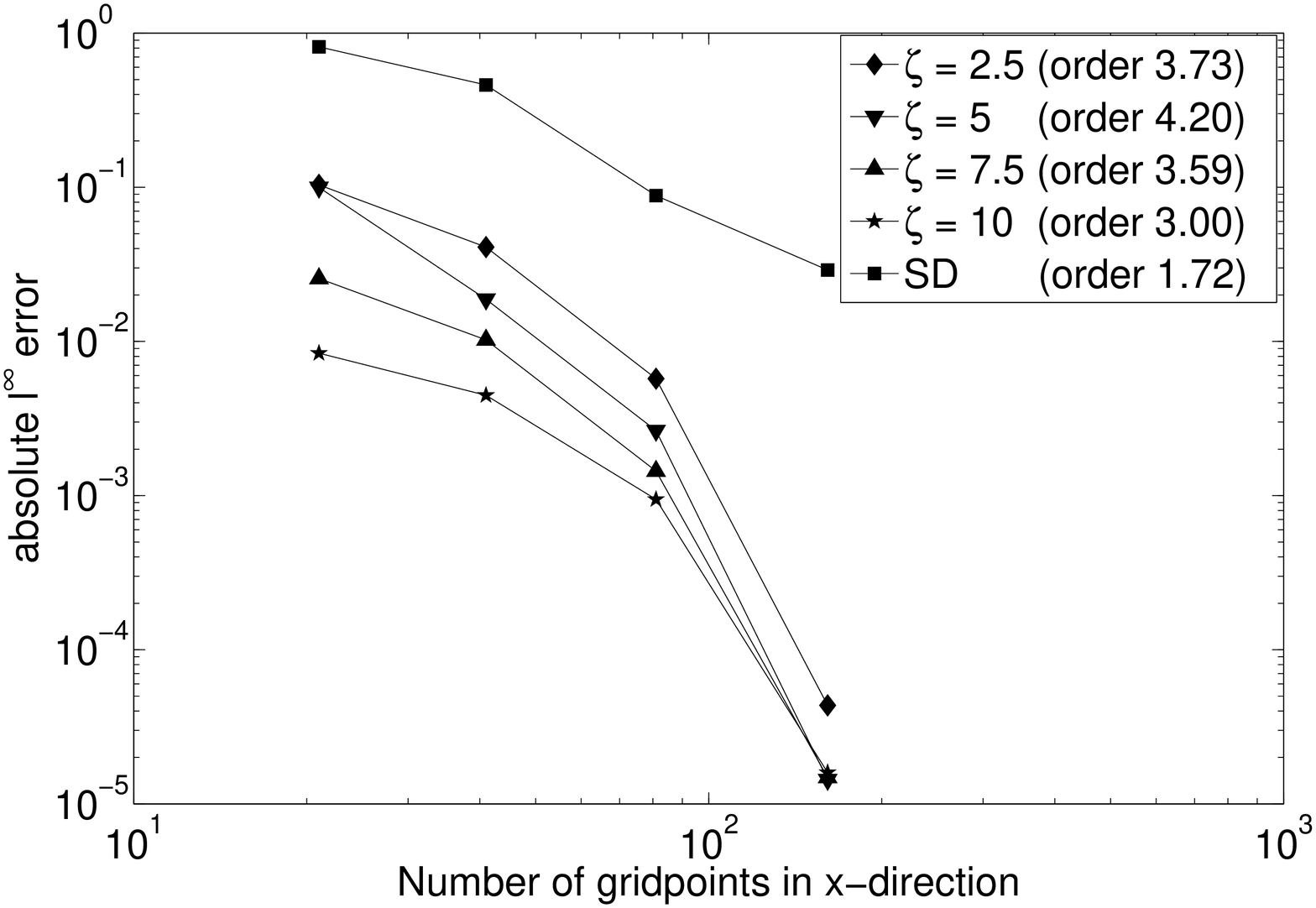}
      \captionof{figure}{Absolute $l^{\infty}$-error Heston model \\$\rho=0$}
      \label{fig:loglog_plot_L_infinity_error_Heston_Hull_White}
    \end{minipage}
\end{center} 
Figures \ref{fig:loglog_plot_L_2_error_Heston_Hull_White} and
\ref{fig:loglog_plot_L_infinity_error_Heston_Hull_White} show log-log
plots of the relative $l^2$- and  $l^{\infty}$-error of the approximations with
respect to the reference solution in the Heston-Hull-White model
($\rho=0$) for a European Put option for different values for the
number of grid points and with different zooms. In this way the influence
of the zoom can be observed. The theoretical consistency order in this
casel is four. Looking at the relative $l^2$-error we observe that the
numerical convergence orders
vary from $3.75$ to $4.29$, which agrees very well with the theoretical
order for all zooms. We can also see that the convergence order rises
until $\zeta=5$ and then declines again, so $\zeta \approx 5$ seems to be the
best choice. The lowest relative $l^2$-error is always obtained when using
$\zeta=10$. 

The more useful error in practice is probably the  $l^{\infty}$-error, as it shows the highest difference between the
reference solution and the approximation. When looking at Figure~\ref{fig:loglog_plot_L_infinity_error_Heston_Hull_White} we see
that the  $l^{\infty}$-error and the $l^2$-error have a very similar
behaviour. The convergence orders vary
from $3.00$ to $4.20,$ again having the best order for
$\zeta \approx 5$. When
using the finest grid the error for $\zeta=5$ and $\zeta=10$ are
almost identical, but with rougher grids the error with $\zeta=10$ is
again clearly the lowest. For both error plots we observe that the
zoom has its biggest impact when looking at a rough grid, because
the error then decreases significantly with an increasing zoom. The HOC discretisations have significantly lower error values and higher convergence orders when comparing them to the standard discretisation. Overall,
choosing $\zeta \approx 5$ for the Heston-Hull-White model ($\rho=0$) seems to be
the best choice with respect to the convergence order.  
\begin{center}
    \begin{minipage}[t]{.49\linewidth}
      \includegraphics[width=8cm,height=5.5cm]{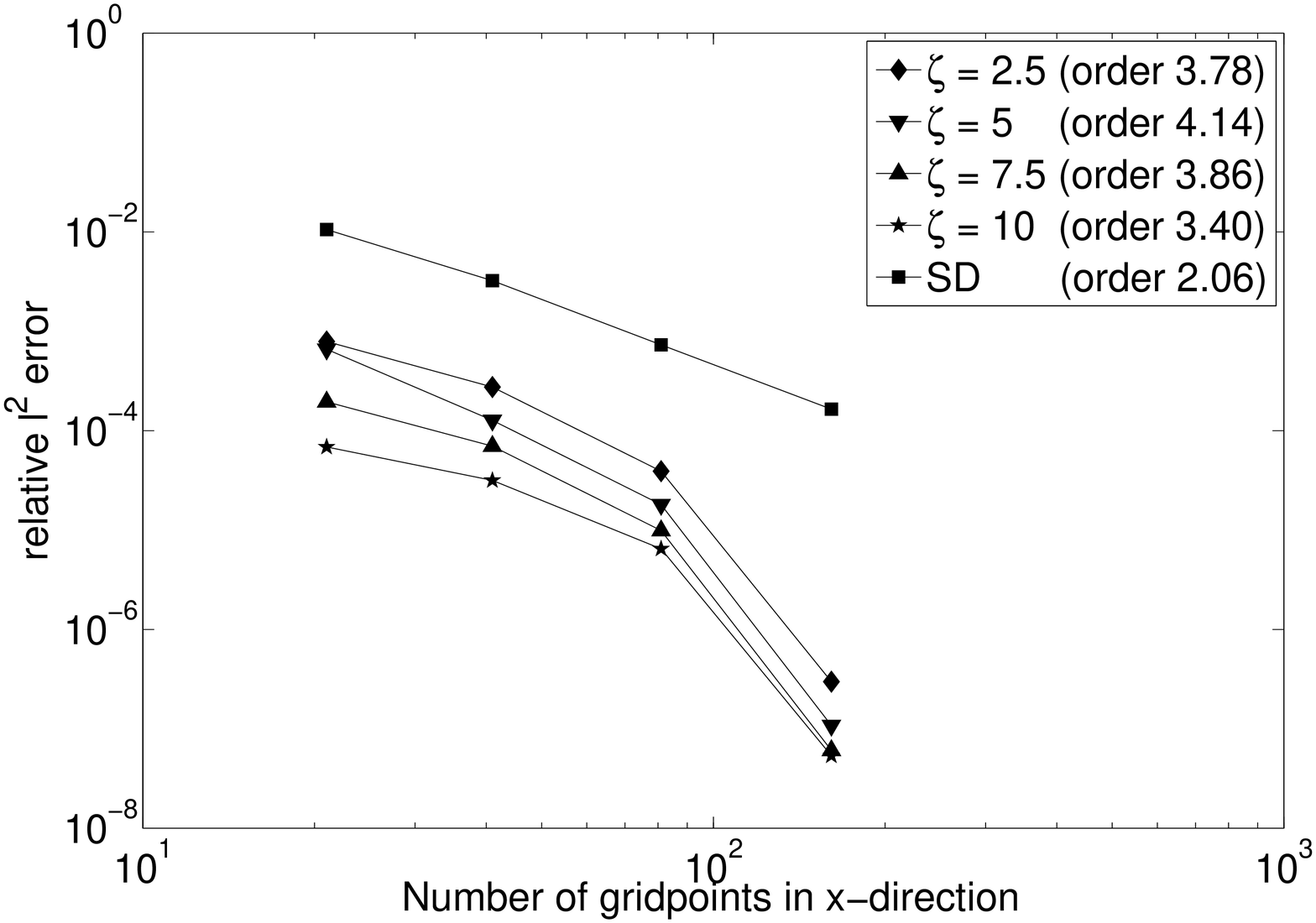}
      \captionof{figure}{Relative $l^2$-error Heston model \\$\rho=-0.1$}
      \label{fig:loglog_plot_L_2_error_Heston_rho=_minus_0_komma_1}
    \end{minipage}%
    \begin{minipage}{.02\linewidth}
    ~
    \end{minipage}%
    \begin{minipage}[t]{.49\linewidth}
      \includegraphics[width=8cm,height=5.5cm]{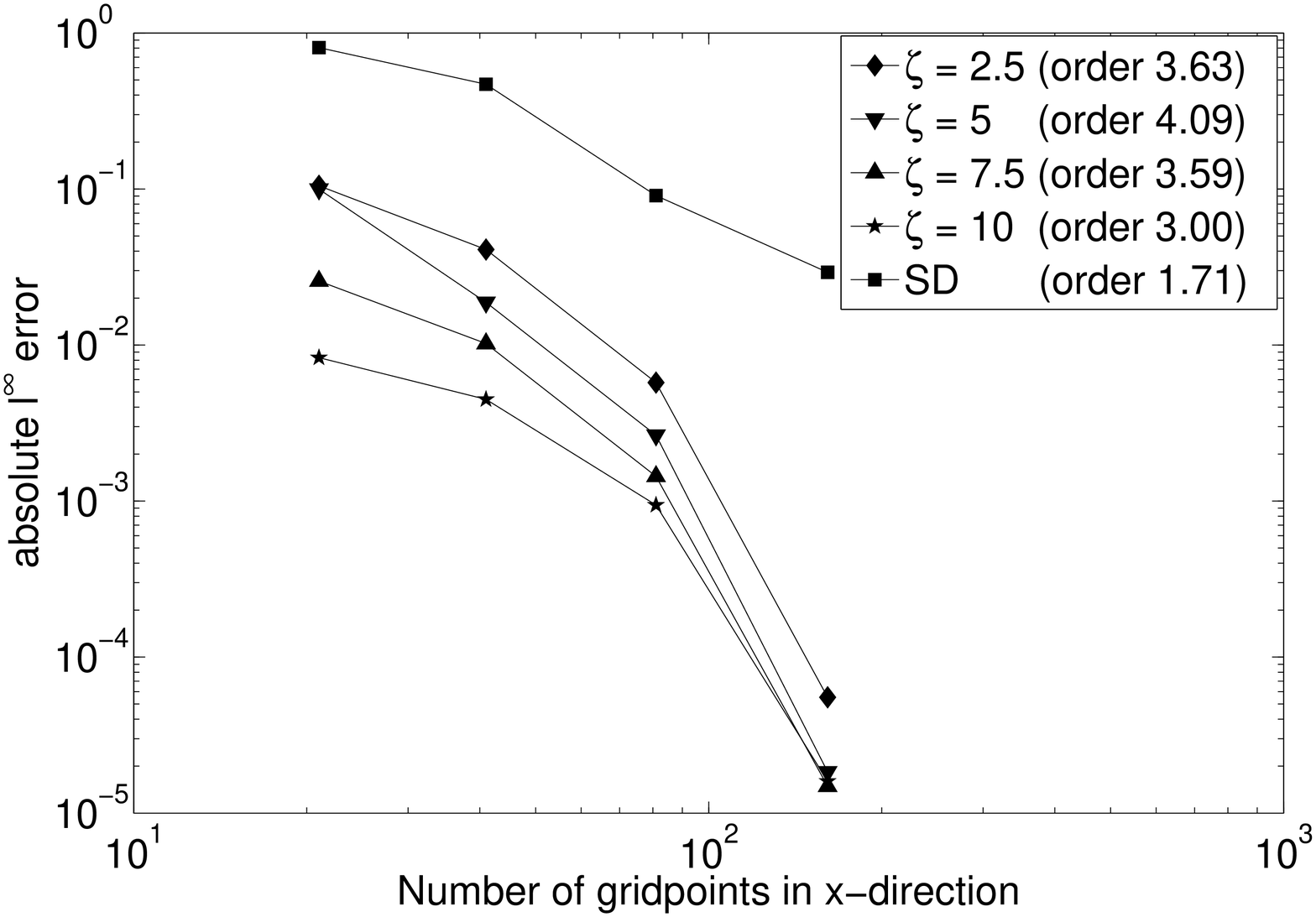}
      \captionof{figure}{Absolute $l^{\infty}$-error Heston model \\$\rho=-0.1$}
      \label{fig:loglog_plot_L_infinity_error_Heston_rho=_minus_0_komma_1}
    \end{minipage}
\end{center} 
In Figures \ref{fig:loglog_plot_L_2_error_Heston_rho=_minus_0_komma_1}
and \ref{fig:loglog_plot_L_infinity_error_Heston_rho=_minus_0_komma_1}
we plot the relative $l^2$- and  $l^{\infty}$-error for a European Put option in
the Heston model with $\rho=-0.1$. This means that the theoretical
consistency order is only two, see equation \eqref{Version3}. We
observe in
Figure~\ref{fig:loglog_plot_L_2_error_Heston_rho=_minus_0_komma_1}
that the relative $l^2$-error varies from $3.40$ to $4.14$. These values
are far above the theoretical consistency order. In fact, using the
Version 3 discretisation scheme we obtain a convergence order close to
the order using the Heston-Hull-White model. 
The order of the relative $l^2$-error is again rising
until $\zeta=5$ and declining afterwards, but has its lowest values
when using $\zeta=10$. The  $l^{\infty}$-error in Figure~\ref{fig:loglog_plot_L_infinity_error_Heston_rho=_minus_0_komma_1}
behaves similar to the  $l^{\infty}$-error in the Heston-Hull-White
model. Here the convergence order values vary from $3.00$ to
$4.09$, having its highest value for $\zeta=5$. With the finest grid
the difference of the error when using $\zeta=10$ and using $\zeta=5$
is again very slim. The biggest impact of increasing the zoom in
either error plot can be again seen when having a rough grid, because
then increasing the zooming leads to significantly lower
errors. Similar as in the Heston-Hull-White model the convergence
order results are the best when choosing $\zeta=5$. For both errors we can again see that the essentially high-order compact discretisations have significantly lower error values and higher convergence orders than the standard discretisation.
\begin{center}
    \begin{minipage}[t]{.49\linewidth}
      \includegraphics[width=8cm,height=5.5cm]{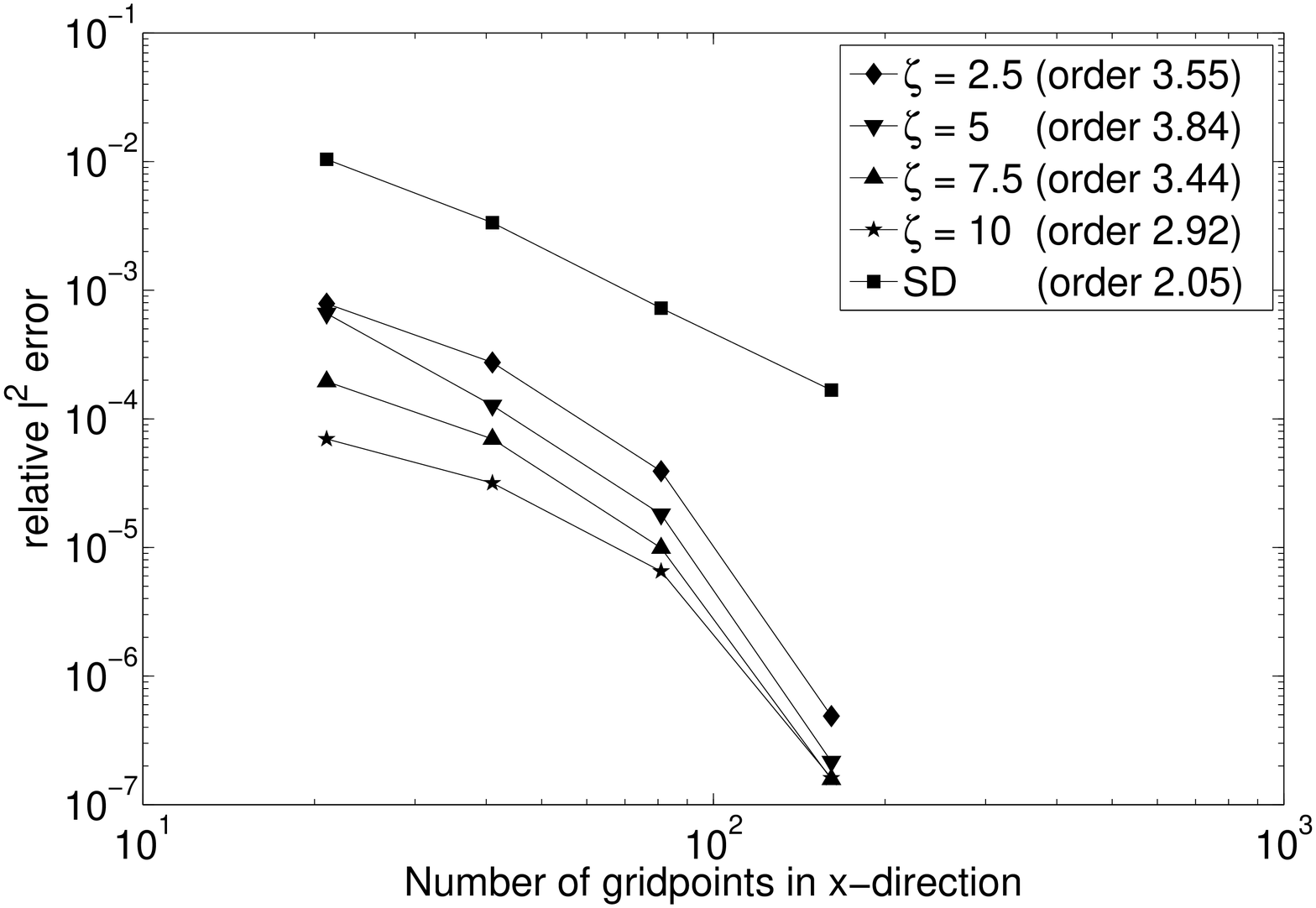}
	  \captionof{figure}{Relative $l^2$-error Heston model \\$\rho=-0.4$}
	  \label{fig:loglog_plot_L_2_error_Heston_rho=_minus_0_komma_4}
    \end{minipage}%
    \begin{minipage}{.02\linewidth}
    ~
    \end{minipage}%
    \begin{minipage}[t]{.49\linewidth}
      \includegraphics[width=8cm,height=5.5cm]{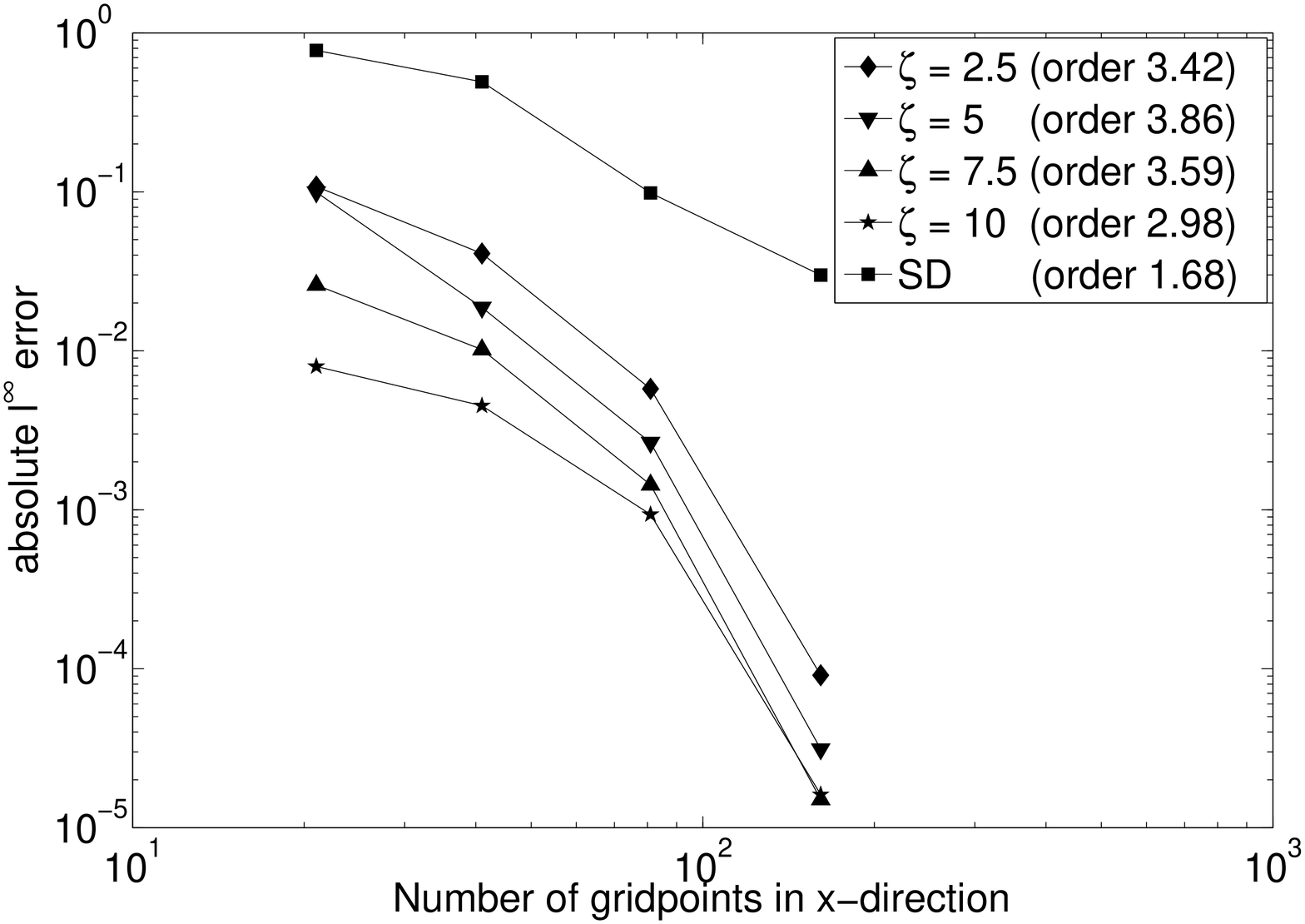}
      \captionof{figure}{Absolute $l^{\infty}$-error Heston model \\$\rho=-0.4$}
      \label{fig:loglog_plot_L_infinity_error_Heston_rho=_minus_0_komma_4}
    \end{minipage}
\end{center} 
Figures \ref{fig:loglog_plot_L_2_error_Heston_rho=_minus_0_komma_4}
and \ref{fig:loglog_plot_L_infinity_error_Heston_rho=_minus_0_komma_4}
show the relative $l^2$- and  $l^{\infty}$-error for an European Put option in
the Heston model with $\rho=-0.4$. The theoretical consistency orders
of the errors are again two. In
Figure~\ref{fig:loglog_plot_L_2_error_Heston_rho=_minus_0_komma_4} we
can see that the convergence order for the relative $l^2$-error varies from $2.92$ to
$3.84$, which is again significantly higher than the theoretical
order. The convergence order deteriorates slightly for smaller values of $\rho$
but is still an order better than for the standard discretisation. 
As expected the best convergence order, which is still very
close to four, will be achieved when using $\zeta=5$. From
Figure~\ref{fig:loglog_plot_L_infinity_error_Heston_rho=_minus_0_komma_4}
we find that for the  $l^{\infty}$-error the convergence order gets
lower with lowering the value of $\rho$. The convergence orders vary
from $2.98$ to $3.86$, where $\zeta=5$ leads again to the highest
value, which is still close to four and thus highly above the
theoretical value of the consistency error order. As in the two previous
cases the zoom has his highest strengths for the relative $l^2$-error as well
as for the  $l^{\infty}$-error when using a very rough grid. For both the relative $l^2$-error and the  $l^{\infty}$-error we can again see that the essentially high-order compact schemes have significantly lower error values and higher convergence orders than the standard discretisation.

With the Figures \ref{fig:loglog_plot_L_2_error_Heston_Hull_White} to \ref{fig:loglog_plot_L_infinity_error_Heston_rho=_minus_0_komma_4} we recover the numerical observation given in Section~\ref{sec:derivation_of_schemes} and can confirm that Version 3 leads to a high-order compact scheme.

For all the discussed European Put options the best results for
the convergence order is obtained when using $\zeta=5$. This value seems to give a
good balance between the error around $K$ and the other regions for
the zoom. Even though the scheme has a
theoretical consistency order equal to four only for the Heston-Hull-White
model ($\rho=0$), the application showed, that we achieve
a numerical convergence order close to four for the Heston model with
$\rho \neq 0$ as well. 

We now consider the case of European Power Put options in the Heston
model. The only difference to a plain vanilla European Put is, that
the final condition is taken to the power $p$,
see \eqref{Final_Condition_Original_Problem}, which yields to
\eqref{Initial_condition_transformed_problem} after transformation. The grid was shifted in
a similar manner as above, avoiding $x_k$ as a grid point.  
\begin{center}
    \begin{minipage}[t]{.49\linewidth}
      \includegraphics[width=8cm,height=5.5cm]{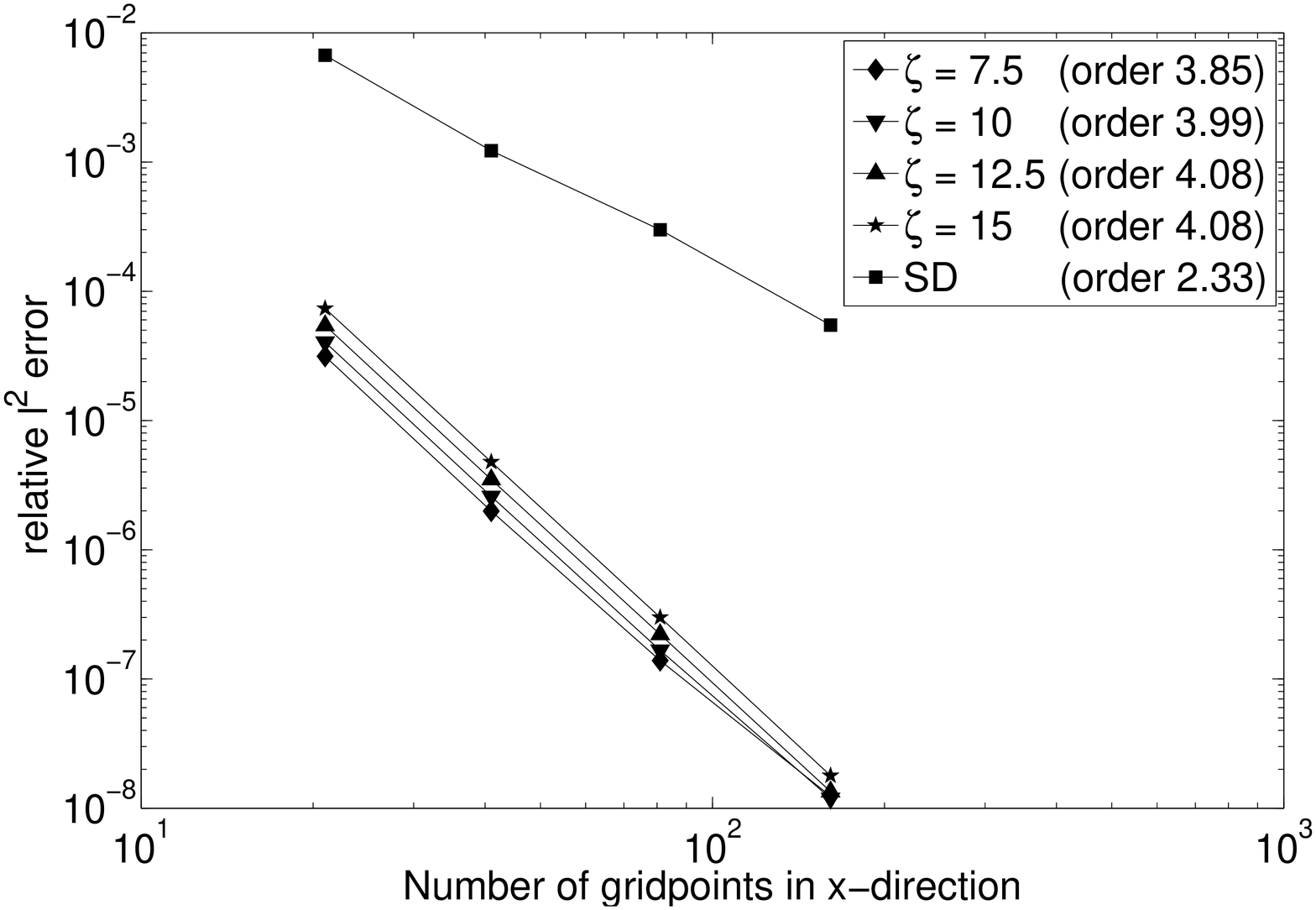}
	  \captionof{figure}{Relative $l^2$-error Power Option Heston model $\rho  = 0$, $p=2$}
	  \label{fig:loglog_plot_L_2_error_Power_Option_Power_2_Heston_Hull_White}
    \end{minipage}%
    \begin{minipage}{.02\linewidth}
    ~
    \end{minipage}%
    \begin{minipage}[t]{.49\linewidth}
      \includegraphics[width=8cm,height=5.5cm]{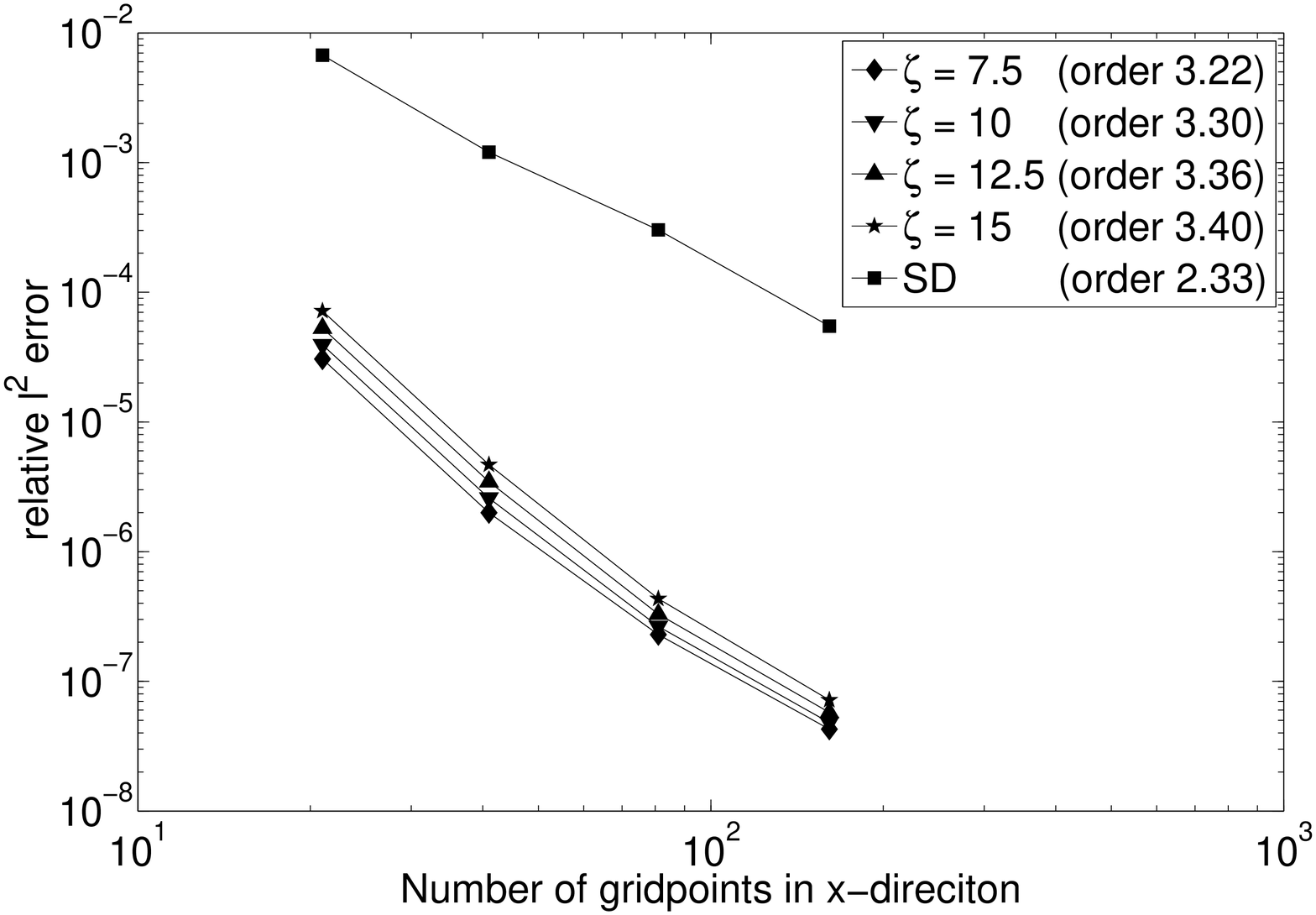}
      \captionof{figure}{Relative $l^{2}$-error Power Option Heston model  $\rho=-0.4$, $p=2$}
      \label{fig:loglog_plot_L_infinity_error_Power_Option_Power_2_Heston_rho=_minus_0_komma_4}
    \end{minipage}
\end{center} 
It can be clearly seen that in Figures
\ref{fig:loglog_plot_L_2_error_Power_Option_Power_2_Heston_Hull_White}
and
\ref{fig:loglog_plot_L_infinity_error_Power_Option_Power_2_Heston_rho=_minus_0_komma_4},
 denoted to the relative $l^2$-error in the cases $\rho = 0$ and $\rho = -0.4 $ when $p=2$, the lines in the log-log plots are much closer to straight lines than in the cases of the vanilla Put options with $p=1$, which can be explained
with the initial condition of the transformed problem being
smoother. The convergence orders of the relative $l^2$-errors range from
$3.85$ to $ 4.08$ for the Heston-Hull-White ($\rho=0$) Power Put with power $p=2$
and from $3.22 $ to $ 3.40$ for the Power Put in the Heston model
with $\rho =-0.4$, where the orders are increasing with increasing
zoom strength. The differences of about $0.6$ between
the orders in the Heston model with $\rho=0$ and $\rho =-0.4$ is not
very large considering the difference of the theoretical orders. So we can again see that the convergence order for $\rho= - 0.4$ is far beyond its theoretical order of two. We can see that the HOC schemes for $\rho = 0$ as well as the essentially high-order compact discretisations for $\rho = - 0.4$ outperform the standard discretisation in terms of error values and convergence orders significantly.
\begin{center}
    \begin{minipage}[t]{.49\linewidth}
      \includegraphics[width=8cm,height=5.5cm]{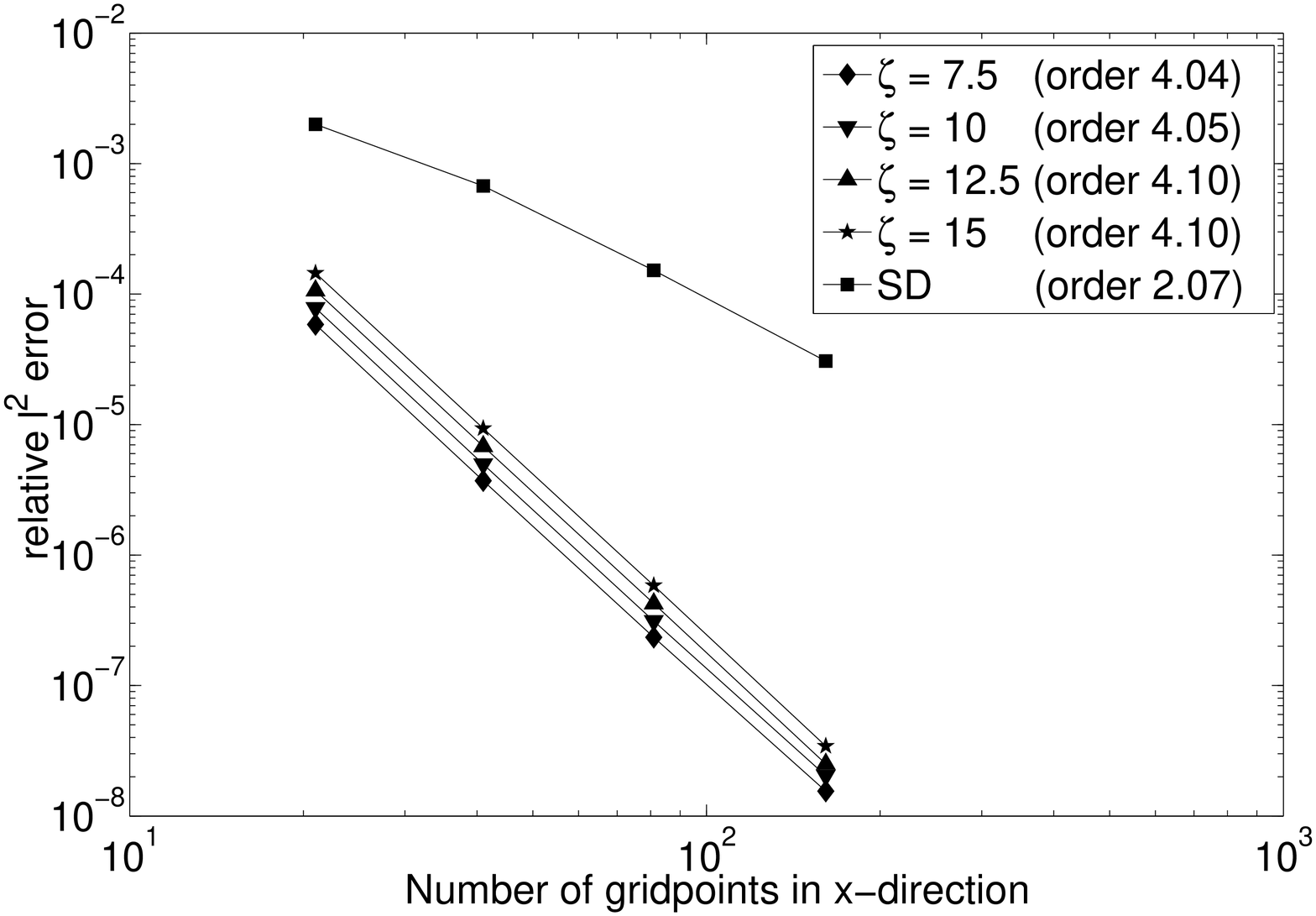}
	  \captionof{figure}{Relative $l^2$-error Power Option Heston model $\rho = 0$, $p=3$}
	  \label{fig:loglog_plot_L_2_error_Power_Option_Power_3_Heston_Hull_White}
    \end{minipage}%
    \begin{minipage}{.02\linewidth}
    ~
    \end{minipage}%
    \begin{minipage}[t]{.49\linewidth}
      \includegraphics[width=8cm,height=5.5cm]{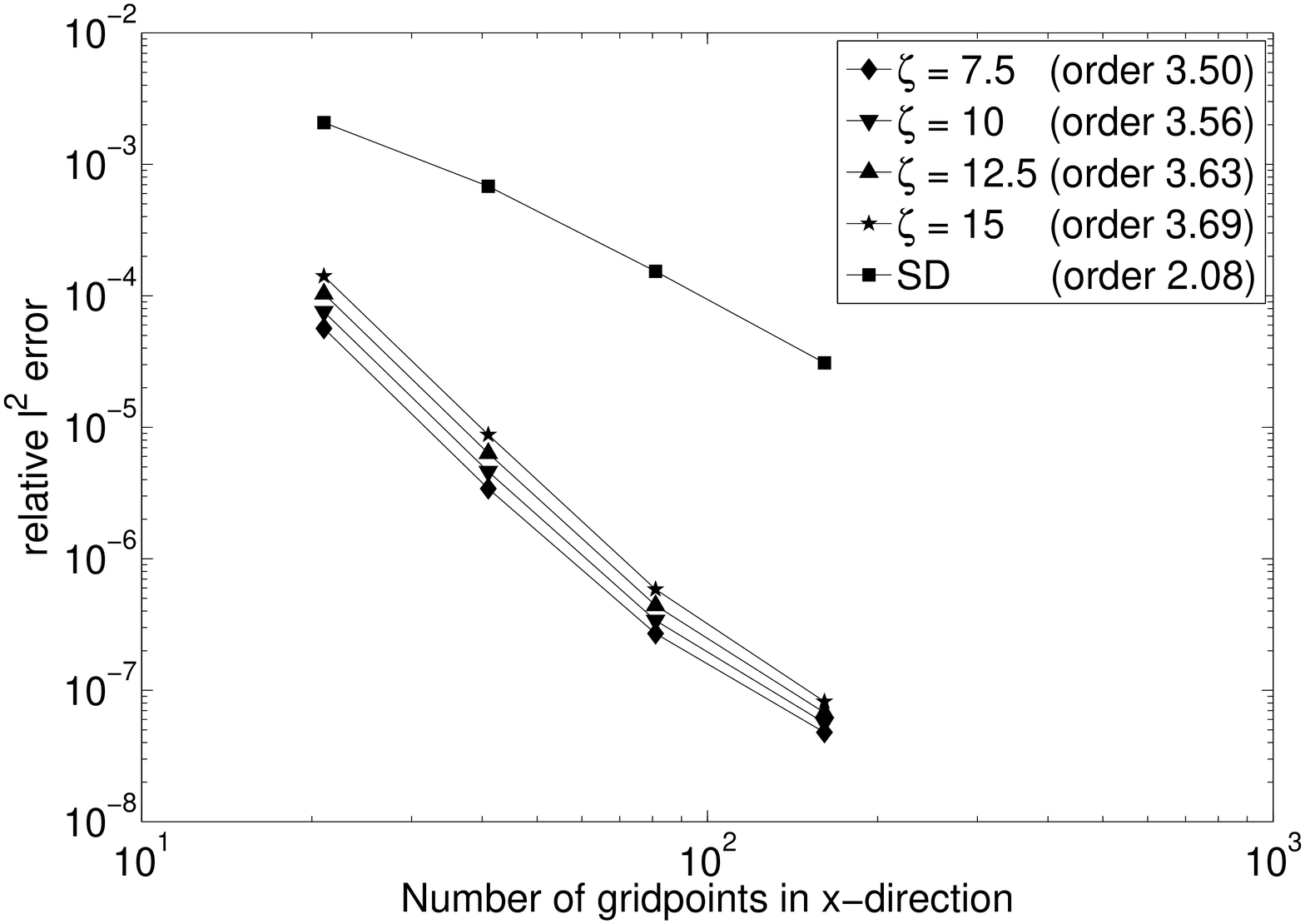}
      \captionof{figure}{Relative $l^{2}$-error Power Option Heston model $\rho=-0.4$, $p=3$}
      \label{fig:loglog_plot_L_infinity_error_Power_Option_Power_3_Heston_rho=_minus_0_komma_4}
    \end{minipage}
\end{center} 
In Figures
\ref{fig:loglog_plot_L_2_error_Power_Option_Power_3_Heston_Hull_White}
and
\ref{fig:loglog_plot_L_infinity_error_Power_Option_Power_3_Heston_rho=_minus_0_komma_4}
we can see the convergence orders
in the Heston-Hull-White model ($\rho=0$) and the Heston model with $\rho= -0.4$ when $p=3$. The differences between the plots
are not as big as the theoretical consistency error order may indicate. Even though
in the Heston model with $\rho= -0.4$ the scheme has a theoretical
consistency error of order two, it produces a convergence order from $3.50$ to $3.69$
depending on the zoom strength $\zeta$, whereas the orders in the
Heston-Hull-White model with $\rho=0$, where we have a theoretical
consistency order of four, vary from $4.04$ to $4.10$. In both situations the standard discretisation is outperformed on behalf of convergence order and error values.

\subsection{Numerical stability study}
\label{sec:stabstudy}

In the particular case of a uniform grid, i.e.\
$\varphi(x)=x,$ the scheme
developed here reduces to the high-order compact scheme presented
in \cite{DuFo12}, where unconditional (von Neumann) stability is
proved for $\rho=0$. An additional stability analysis performed in
\cite{DuFo12p} suggests that the scheme is also unconditionally stable
for general choice of parameters. For the present scheme on a
non-uniform grid, a similar von Neumann analysis, analytical or
numerical, appears to be out of reach as the expression for the
amplification factor is formidable and consists of high-order
polynomials in a two-digit number of variables. 
To validate the stability of the scheme for general
parameters, we therefore perform additional numerical stability
tests. We remark that in our numerical experiments we
observe a stable behaviour throughout.

We compute numerical solutions for varying values
of the parabolic mesh ratio $c=\Delta\tau/h^2$ and the mesh width $h.$ Plotting
the associated relative $l^2$-norm errors in the plane should allow us to detect
stability restrictions depending on $c$ or oscillations that occur
for high cell Reynolds number (large $h$). This approach for a numerical
stability study was also used in \cite{DuFo12,DuFoJu03}.

We show results for the European Put option in the Heston Model only, 
since the Power Puts only differ in the initial conditions and give
similar results. 
For our stability
plots we use $c = {k}/{10}$ with $k=1, \ldots , 10$, and a descending
sequence of grid points in $x$-direction, starting with six grid points
(since $x\in [0,1]$ it follows $h \leq 0.2$), and doubling the
number of points (halving $h$) in each step.
The zoom parameter $\zeta = 5$ is used. 
\begin{center}
    \begin{minipage}[t]{.49\linewidth}
      \includegraphics[width=7cm,height=5cm]{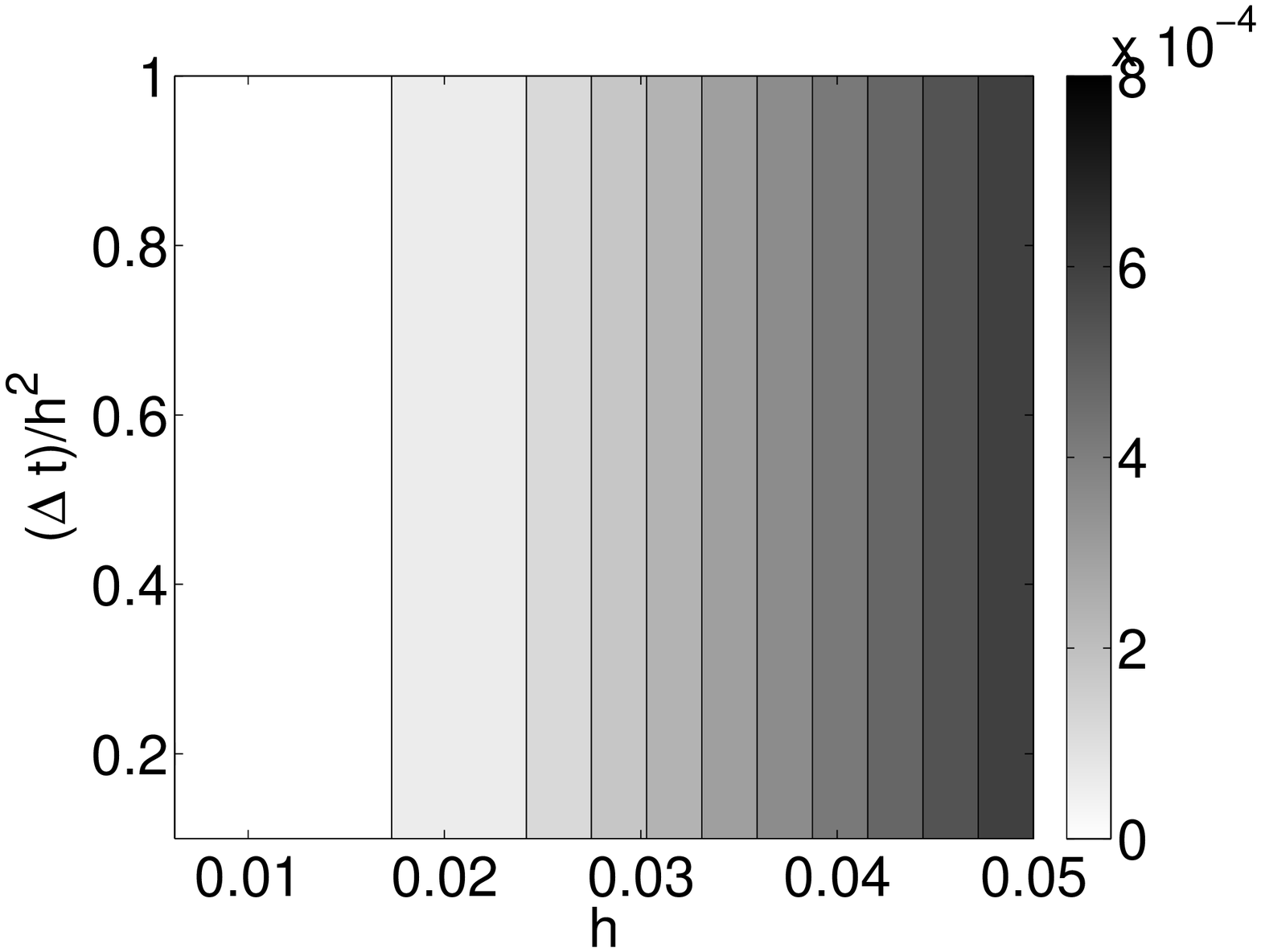}
	  \captionof{figure}{Stability plot of the relative $l^2$-error for $\rho=0$}
	  \label{fig:Stability_Plot_roh=0_L_2_Error}
    \end{minipage}%
    \begin{minipage}{.02\linewidth}
    ~
    \end{minipage}%
    \begin{minipage}[t]{.49\linewidth}
      \includegraphics[width=7cm,height=5cm]{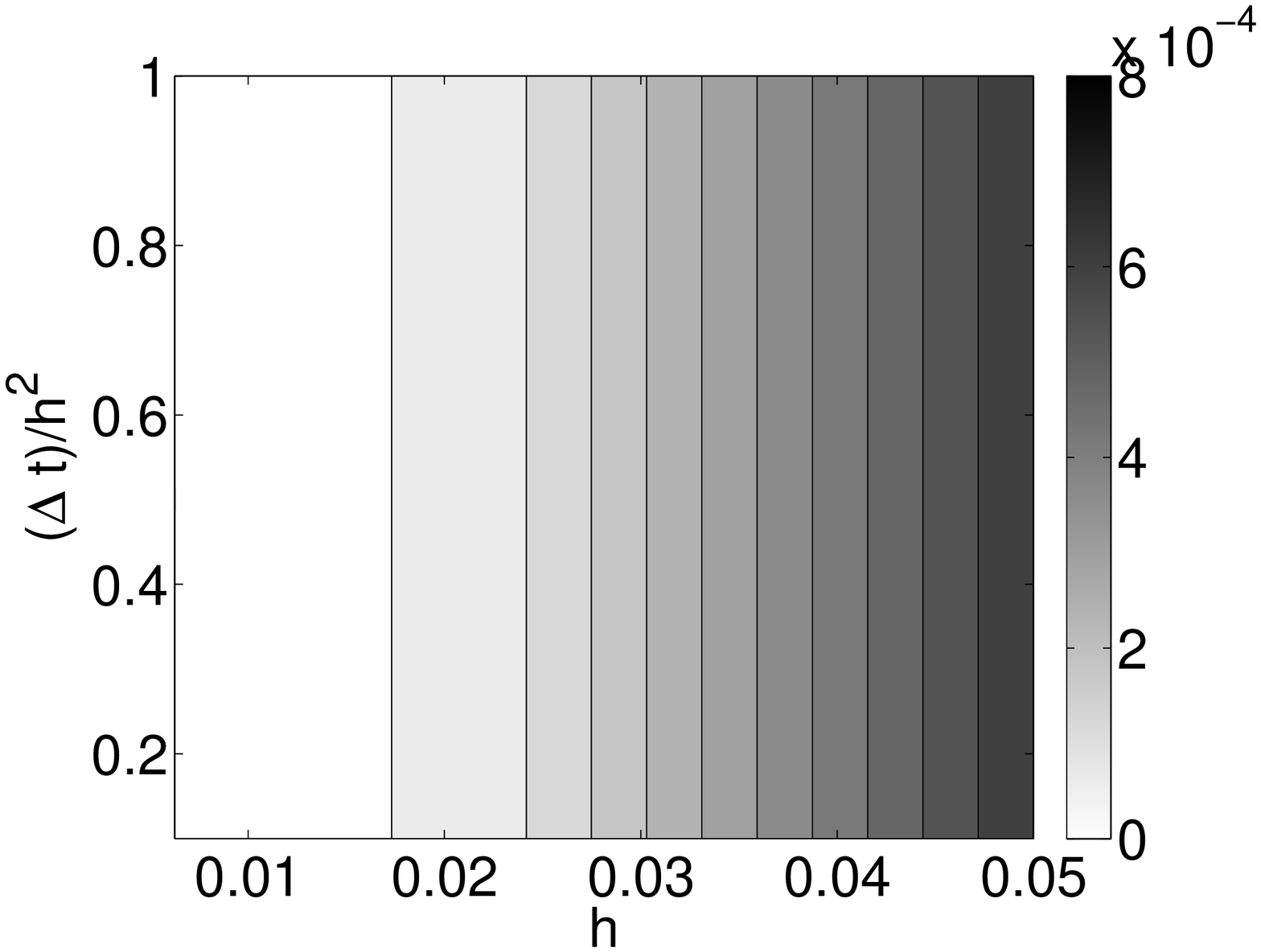}
      \captionof{figure}{Stability plot of the relative $l^2$-error for $\rho=-0.4$}
      \label{fig:Stability_Plot_roh=_minus_0_komma_4_L_2_Error}
    \end{minipage}
\end{center}
Figures \ref{fig:Stability_Plot_roh=0_L_2_Error} and
\ref{fig:Stability_Plot_roh=_minus_0_komma_4_L_2_Error} show the
stability plots for the Heston-Hull-White model ($\rho=0$) and for the Heston
model with $\rho = -0.4$. We observe that the influence of the
parabolic mesh ratio $c$ on the
relative $l^2$-error is only marginal and the relative error does not exceed $8\times 10^{-4}$ as a
value for both stability plots. We can infer that there
does not seem to be a stability condition on $c$ for either
situation. For increasing values of $h,$ which also result
in a higher cell Reynolds number, the error grows gradually, and no
oscillations in the numerical solutions occur.
The stability plot for the Heston model with $\rho = -0.1$
looks similar (not shown here) and does not indicate any conditions on
$c$ or $h$ either.

\section{Conclusion}
\label{sec:conc}

We have presented new high-order compact finite difference schemes
for option pricing under stochastic volatility on non-uniform grids.
The resulting schemes are fourth-order
accurate in space and second-order accurate in time for vanishing
correlation. In our numerical convergence study we obtain high-order numerical
convergence also for non-zero correlation and 
non-smooth payoffs which are typical in option pricing.
In all numerical experiments a comparative standard second-order discretisation is
significantly outperformed.
 We have conducted a numerical stability study which seems to indicate
unconditional stability of the scheme. 
In our numerical experiments we
observe a stable behaviour for all choices of parameters.

It would be interesting to consider extensions of this scheme to
the American option pricing problem, where early exercise of the option is
possible. In this case, one has to solve a free
boundary problem. It can be written as a linear complementarity problem which could be
discretised using the schemes given here.
To retain the high-order convergence one would need to combine the high-order
discretisation with a high-order resolution of the free boundary.
This extension is beyond the scope of the present paper, and we
leave it for future research.


%% file: HestonVersion2_as_Appendix.tex
\section{Coefficients for Version 2 and Version 4}

In this section we give the coefficients of the semi-discrete schemes
for Version 2 and Version 4. We do not include the coefficients for
Version 1 as this version always resulted into a second-order
numerical convergence error in the numerical study.
\subsection{Coefficients for Version 2}
When discretising equation \eqref{Version2} with the
central difference operator in $x$- and in $y$-direction, we get the
following coefficients for the Version 2 scheme
\begin{eqnarray}
\notag \begin{array}{rcl}
\hat{K}_{i-1,j \pm 1} & = &
 \,{\frac {vy{\varphi^{2} _{{x}}}\varphi _{{{\it xx}}}}{12h}}
 \pm \,{\frac {y{\varphi^{3} _{{x}}}\kappa}{12h}}
 \pm \,{\frac {{\varphi^{4} _{{x}}}\kappa\,\theta\,r}{12{v}^{2}y}}
- \,{\frac {vy\varphi _{{x}}}{12{h}^{2}}}
- \,{\frac {vy\varphi _{{{\it xx}}}}{24h}}
 \pm \frac{y{\varphi^{4} _{{x}}}\kappa}{24}
+ \,{\frac {{\varphi^{2} _{{x}}}r}{12h}}
 \pm \,{\frac {\varphi _{{{\it xx}}}\kappa\,{\varphi^{2} _{{x}}}\theta}{24v}}
- \,{\frac {vy{\varphi^{2} _{{x}}}}{24h}} \\
\\
&&
 \pm \,{\frac {{\varphi^{4} _{{x}}}r}{24y}}
 \mp \,{\frac {{\varphi^{2} _{{x}}}r}{12y}}
 \mp \frac{y\varphi _{{{\it xx}}}\kappa\,{\varphi^{2} _{{x}}}}{24}
 \mp \,{\frac {\kappa\,{\varphi^{3} _{{x}}}\theta}{12hv}}
 \mp \,{\frac {{\varphi^{4} _{{x}}}\kappa\,\theta}{24v}}
 \mp \,{\frac {{\varphi^{4} _{{x}}}\kappa\,r}{12v}} \\
\\
&&
+\rho\left[
 \mp \frac{{\varphi^{2} _{{x}}} \left( \frac{vy}{2}-r \right) \varphi _{{{\it xx}}}}{24}
 \pm \frac{vy\varphi _{{x}}\varphi _{{{\it xxx}}}}{48}
 \mp \,{\frac {v{\varphi^{2} _{{x}}}}{12y}}
 \mp \frac{vy{\varphi^{2} _{{{\it xx}}}}}{48}
 \pm \,{\frac {v{\varphi^{4} _{{x}}}}{24y}}
 \mp \frac{{\varphi^{4} _{{x}}}\kappa}{24}
 \pm \,{\frac {vy{\varphi^{2} _{{x}}}}{4{h}^{2}}}
 \pm \,{\frac {{\varphi^{3} _{{x}}} \left( \frac{vy}{2}-r \right) }{6h}}  \right.\\
\\
&& \left.
 \pm \,{\frac {vy\varphi _{{x}}\varphi _{{{\it xx}}}}{12h}}
+ \,{\frac {{\varphi^{4} _{{x}}}\kappa\, \left( \theta-vy \right) }{6hv}}
\right] 
+{\rho}^{2} \left[
- \,{\frac {vy{\varphi^{3} _{{x}}}}{6{h}^{2}}}
 \mp \frac{v{\varphi^{2} _{{x}}}\varphi _{{{\it xx}}}}{8}
- \,{\frac {vy{\varphi^{2} _{{x}}}\varphi _{{{\it xx}}}}{12h}}
\right] ,
\\
\\
\hat{K}_{i+1,j \pm 1} & = & 
- \hat{K}_{i-1,j \pm 1}  
 \pm \,{\frac {y{\varphi^{3} _{{x}}}\kappa}{6h}}
 - \,{\frac {vy\varphi _{{x}}}{6{h}^{2}}}
 \mp \,{\frac {\kappa\,{\varphi^{3} _{{x}}}\theta}{6hv}}
 \pm \,{\frac {\rho {\varphi^{3} _{{x}}} \left( \frac{vy}{2}-r \right) }{3h}} 
 \pm \,{\frac {\rho vy\varphi _{{x}}\varphi _{{{\it xx}}}}{6h}}
-  \,{\frac {{\rho}^{2}vy{\varphi^{3} _{{x}}}}{3{h}^{2}}} , 
\\
\\
\hat{K}_{i,j \pm 1} & = &
- \,{\frac {vy{\varphi^{3} _{{x}}}}{2{h}^{2}}}
 \pm \,{\frac {y{\varphi^{3} _{{x}}}\kappa}{3h}}
 \mp \,{\frac {\kappa\,{\varphi^{3} _{{x}}}\theta}{3hv}}
+ \,{\frac {v{\varphi^{3} _{{x}}}}{6y}}
 \mp \frac{y\varphi _{{x}}h{\varphi^{2} _{{{\it xx}}}}\kappa}{8}
 \mp \frac{y{\varphi^{3} _{{x}}}h\kappa\,\varphi _{{{\it xx}}}}{8}
- \,{\frac {{\varphi^{5} _{{x}}}\kappa\,\theta}{4vy}}
- \,{\frac {{\varphi^{5} _{{x}}}{\kappa}^{2}{\theta}^{2}}{6{v}^{3}y}}
 \\
\\
&&
- \frac{vy{\varphi^{2} _{{x}}}\varphi _{{{\it xxx}}}}{8} 
 \mp \,{\frac {h{\varphi^{3} _{{x}}}\kappa}{6y}}
 \mp \,{\frac {h{\varphi^{3} _{{x}}}\kappa\,\varphi _{{{\it xx}}}\theta\,r}{4{v}^{2}y}}
- \frac{{\varphi^{3} _{{x}}}\varphi _{{{\it xx}}}r}{4}
- \,{\frac {v{\varphi^{5} _{{x}}}}{12y}}
- \,{\frac {y{\varphi^{5} _{{x}}}{\kappa}^{2}}{6v}}
 \mp \,{\frac {{\varphi^{5} _{{x}}}h{\kappa}^{2}}{12v}}
 \\
\\
&&
+ \frac{vy\varphi _{{x}}{\varphi _{{{\it xx}}}}^{2}}{8}
+ \frac{vy{\varphi^{3} _{{x}}}\varphi _{{{\it xx}}}}{8}
 \pm \,{\frac {{\varphi^{5} _{{x}}}h\kappa}{12y}}
+ \,{\frac {{\varphi^{5} _{{x}}}{\kappa}^{2}\theta}{3{v}^{2}}}
 \pm \frac{hy{\varphi^{2} _{{x}}}\kappa\,\varphi _{{{\it xxx}}}}{8}
+ \,{\frac {\kappa\,{\varphi^{3} _{{x}}}\theta}{3vy}}
 \pm \,{\frac {h{\varphi^{3} _{{x}}}\kappa\,\varphi _{{{\it xx}}}\theta}{8v}}
 \\
\\
&&
 \pm \,{\frac {h{\varphi^{3} _{{x}}}\kappa\,\varphi _{{{\it xx}}}r}{4v}}
 \pm \,{\frac {\varphi _{{x}}h{\varphi^{2} _{{{\it xx}}}}\kappa\,\theta}{8v}}
 \pm \,{\frac {{\varphi^{5} _{{x}}}h{\kappa}^{2}\theta}{12{v}^{2}y}} 
- \mp \,{\frac {h{\varphi^{2} _{{x}}}\kappa\,\varphi _{{{\it xxx}}}\theta}{8v}}
+ \,{\frac {vy\varphi _{{x}}}{6{h}^{2}}}
+ \frac{{\varphi^{5} _{{x}}}\kappa}{12}
\\
\\
&&
\rho\left[
 \frac{v{\varphi^{3} _{{x}}}\varphi _{{{\it xx}}}}{4}
 \pm \,{\frac {h{\varphi^{3} _{{x}}}\kappa\, \left( \theta-vy \right) \varphi _{{{\it xx}}}}{4vy}}
 \mp \,{\frac {{\varphi^{3} _{{x}}} \left( \frac{vy}{2}-r \right) }{3h}}
 \mp \,{\frac {vy\varphi _{{x}}\varphi _{{{\it xx}}}}{6h}}
\right]
+{\rho}^{2}
 \,{\frac {vy{\varphi^{3} _{{x}}}}{3{h}^{2}}} ,
\\
\\
\hat{K}_{i \pm 1,j} & = &
 \mp \,{\frac {h\varphi _{{{\it xx}}}v}{12y}}
 \mp \frac{h\varphi _{{x}}\varphi _{{{\it xxx}}}r}{24}
 \mp \,{\frac {h{\varphi^{2} _{{x}}}v}{12y}}
 \pm \,{\frac {vy{\varphi^{2} _{{x}}}}{6h}}
 \pm \,{\frac {vy{\varphi^{2} _{{x}}}\varphi _{{{\it xx}}}}{6h}}
+ \frac{{\varphi^{3} _{{x}}}r}{6}
 \pm \frac{h{\varphi^{2} _{{{\it xx}}}}r}{6}
+ \frac{vy\varphi _{{{\it xxx}}}}{24}
- \,{\frac {v{\varphi^{3} _{{x}}}}{12y}} \\
\\
&&
 \pm \,{\frac {h\varphi _{{{\it xx}}}v{\varphi^{2} _{{x}}}}{24y}}
 \mp \frac{hvy{\varphi^{2} _{{x}}}\varphi _{{{\it xx}}}}{24}
- \frac{vy{\varphi^{3} _{{x}}}}{24} 
 \pm \,{\frac {h{\varphi^{2} _{{x}}}\kappa\,\varphi _{{{\it xx}}}\theta}{24vy}}
+ \frac{vy\varphi _{{x}}\varphi _{{{\it xx}}}}{24}
 \pm \frac{hvy\varphi _{{{\it xxxx}}}}{48}
 \mp \frac{hvy{\varphi^{2} _{{{\it xx}}}}}{12}
+ \,{\frac {v\varphi _{{x}}}{6y}}
\\
\\
&&
- \frac{\varphi _{{x}}\varphi _{{{\it xx}}}r}{12} 
+ \frac{\kappa\,{\varphi^{3} _{{x}}}}{12}
 \mp \frac{h{\varphi^{4} _{{x}}}\kappa}{24}
 \pm \,{\frac {vy\varphi _{{{\it xx}}}}{6h}} 
 \pm \,{\frac {{\varphi^{4} _{{x}}}hv}{24y}}
 \mp \frac{h{\varphi^{2} _{{x}}}\kappa\,\varphi _{{{\it xx}}}}{24}
 \pm \frac{h{\varphi^{2} _{{x}}}\varphi _{{{\it xx}}}r}{6}
- \,{\frac {{\varphi^{3} _{{x}}}{r}^{2}}{6vy}} 
 \pm \,{\frac {h{\varphi^{4} _{{x}}}\kappa\,\theta}{24vy}}\\
\\
&&
 \mp \,{\frac {{\varphi^{2} _{{x}}}h\varphi _{{{\it xx}}}{r}^{2}}{6vy}}
- \,{\frac {\kappa\,{\varphi^{3} _{{x}}}\theta}{12vy}}
 \pm \frac{hvy\varphi _{{x}}\varphi _{{{\it xxx}}}}{48} 
 \mp \,{\frac {{\varphi^{2} _{{x}}}r}{3h}}
 \mp \,{\frac {hvy\varphi _{{{\it xx}}}\varphi _{{{\it xxx}}}}{16\varphi _{{x}}}}
- \,{\frac {vy\varphi _{{x}}}{3{h}^{2}}}
+{\rho}^{2}\left[
 \,{\frac {vy{\varphi^{3} _{{x}}}}{3{h}^{2}}}
\mp  \,{\frac {vy{\varphi^{2} _{{x}}}\varphi _{{{\it xx}}}}{6h}}
\right] 
\\
\\
&&
+\rho \left[
 \frac{v{\varphi^{3} _{{x}}}}{12}
+ \frac{v\varphi _{{x}}\varphi _{{{\it xx}}}}{4}
 \mp \frac{h\varphi _{{{\it xx}}}v{\varphi^{2} _{{x}}}}{24}
 \mp \frac{hv{\varphi^{2} _{{{\it xx}}}}}{8}
 \mp \,{\frac {h{\varphi^{2} _{{x}}} \left( \frac{vy}{2}-r \right) \varphi _{{{\it xx}}}}{6y}}
- \,{\frac {{\varphi^{3} _{{x}}} \left( \frac{vy}{2}-r \right) }{6y}}
 \pm \,{\frac {{\varphi^{4} _{{x}}}\kappa\, \left( \theta-vy \right) }{3hv}}
\right] 
\end{array}
\end{eqnarray}
and
\begin{eqnarray}
\notag \begin{array}{rcl}
\hat{K}_{i,j} & = &
- \frac{\kappa\,{\varphi^{3} _{{x}}}}{6}
+ \frac{vy{\varphi^{2} _{{x}}}\varphi _{{{\it xxx}}}}{4}
- \frac{vy\varphi _{{x}}{\varphi^{2} _{{{\it xx}}}}}{4}
- \frac{{\varphi^{5} _{{x}}}\kappa}{6}
+ \,{\frac {v{\varphi^{5} _{{x}}}}{6y}}
- \,{\frac {v{\varphi^{3} _{{x}}}}{6y}}
- \frac{vy\varphi _{{{\it xxx}}}}{12}
+ \,{\frac {2vy\varphi _{{x}}}{3{h}^{2}}}
+ {\frac {vy{\varphi^{3} _{{x}}}}{{h}^{2}}} 
- \,{\frac {2{\varphi^{5} _{{x}}}{\kappa}^{2}\theta}{3{v}^{2}}}\\
\\
&&
+ \frac{{\varphi^{3} _{{x}}}\varphi _{{{\it xx}}}r}{2}
+ \frac{\varphi _{{x}}\varphi _{{{\it xx}}}r}{6}
+ \,{\frac {{\varphi^{5} _{{x}}}\kappa\,\theta}{2vy}}
+ \,{\frac {y{\varphi^{5} _{{x}}}{\kappa}^{2}}{3v}}
- \frac{{\varphi^{3} _{{x}}}r}{3}
- \frac{vy{\varphi^{3} _{{x}}}\varphi _{{{\it xx}}}}{4}
- \frac{vy\varphi _{{x}}\varphi _{{{\it xx}}}}{12}
- \,{\frac {\kappa\,{\varphi^{3} _{{x}}}\theta}{2vy}}
- \,{\frac {v\varphi _{{x}}}{3y}} \\
\\
&&
+ \,{\frac {{\varphi^{5} _{{x}}}{\kappa}^{2}{\theta}^{2}}{3{v}^{3}y}}
+ \frac{vy{\varphi^{3} _{{x}}}}{12}
+ \,{\frac {{\varphi^{3} _{{x}}}{r}^{2}}{3vy}}
+\rho\,[
- \frac{v{\varphi _{{x}}}^{3}\varphi _{{{\it xx}}}}{2}
- \frac{v{\varphi _{{x}}}^{3}}{6}
- \frac{v\varphi _{{x}}\varphi _{{{\it xx}}}}{2}
+ \,{\frac {{\varphi _{{x}}}^{3} \left( \frac{vy}{2}-r \right) }{3y}}
]
-{\rho}^{2}
 \,{\frac {2vy{\varphi _{{x}}}^{3}}{3{h}^{2}}} ,
\end{array}
\end{eqnarray}
where $\hat{K}_{i,j}$ is the coefficient of $U_{i,j}(\tau)$. Defining $\hat{M}_{i,j}$ as the coefficient of $\partial_{\tau}U_{i,j}(\tau)$ we get
\begin{eqnarray}
\notag \begin{array}{rcl}
\hat{M}_{i+1,j \pm 1} & = & \hat{M}_{i-1,j \mp 1} = \pm \frac{\rho\,{\varphi^{4} _{{x}}}}{24} ,
\\
\\
\hat{M}_{i,j\pm 1} & = & 
- \frac{{\varphi^{5} _{{x}}}}{12}
+ \frac{{\varphi^{3} _{{x}}}}{6}
 \mp \,{\frac {{\varphi^{3} _{{x}}}h}{6y}}
 \pm \,{\frac {{\varphi^{5} _{{x}}}h}{12y}}
 \pm \,{\frac {{\varphi^{5} _{{x}}}h\kappa\, \left( \theta-vy \right) }{12{v}^{2}y}} ,
 \\
 \\
\hat{M}_{i \pm 1,j} & = & 
 \frac{{\varphi^{3} _{{x}}}}{12}
 \mp \,{\frac {{\varphi^{4} _{{x}}}h \left( \frac{vy}{2}-r \right) }{12vy}}
 \pm \frac{{\varphi^{2} _{{x}}}h\varphi _{{{\it xx}}}}{8}
 \mp \,{\frac {{\varphi^{4} _{{x}}}h\rho}{12y}} \text{ and }
\\
\\
\hat{M}_{i,j} & = & 
- \,{\frac {{\varphi^{3} _{{x}}}\varphi _{{{\it xx}}}{h}^{2} \left( \frac{vy}{2}-r \right) }{2vy}}
- \frac{\varphi _{{x}}{\varphi^{2} _{{{\it xx}}}}{h}^{2}}{4}
+ \frac{{\varphi^{5} _{{x}}}}{6}
+ \frac{{\varphi^{3} _{{x}}}}{2}
+ \frac{{\varphi^{2} _{{x}}}{h}^{2}\varphi _{{{\it xxx}}}}{4}
- \,{\frac {\rho{\varphi^{3} _{{x}}}\varphi _{{{\it xx}}}{h}^{2}}{2y}}.
\end{array}
\end{eqnarray}
Using these coefficients instead of the ones given in
\eqref{Scheme_Heston_Model_Version_3_K_equation_one} to
\eqref{Scheme_Heston_Model_Version_3_M1_to_M9} in the derivation in
Section~\ref{sec:HOC_schemes_for_parabolic_problems} for the interior
of the grid $G$ as well as the boundaries $y_{\min}$ and $y_{\max}$
yields the Version 2 scheme.

%% file: HestonVersion4_as_Appendix.tex
\subsection{Coefficients for Version 4}
In this part of the appendix we give the coefficients of the Version 4 scheme. When discretising equation \eqref{Version2} with the central difference operator in $x$- and in $y$-direction, we get
\begin{eqnarray}\label{Scheme_Heston_Model_Version_4_K_equation_one}
\notag \begin{array}{rcl}
\hat{K}_{i\pm 1,j} & = & 
\frac {vy{\varphi^{3} _{{x}}}}{12{h}^{2}} 
\mp \frac{h{\varphi^{2} _{{{\it xx}}}} \left( \frac{vy}{2}-r \right)}{6}  
\mp \frac {{\varphi^{4} _{{x}}} \left( \frac{vy}{2}-r \right) }{12h} 
\pm \frac { 5\left( \frac{vy}{2}-r \right) {\varphi^{2} _{{x}}}}{12h} 
\pm \frac{yhv\varphi _{{{\it xxxx}}}}{48} 
\mp \frac {h\varphi _{{{\it xx}}}v}{24y} \\
\\
&& 
- \frac {\varphi _{{x}}\kappa\, \left( \theta-vy \right) }{12vy} 
- \frac {5vy\varphi _{{x}}}{12{h}^{2}} 
\pm \frac {5vy\varphi _{{{\it xx}}}}{24h} 
+ \frac {v\varphi _{{x}}}{12y} 
\mp \frac {{\varphi^{2} _{{x}}}hv}{24y} 
- \frac {{\varphi^{3} _{{x}}} \left( \frac{vy}{2}-r \right) ^{2}}{6vy} 
 + \frac{vy\varphi _{{{\it xxx}}}}{24} \\
\\
&&
\pm \frac{\varphi _{{x}}h \left( \frac{vy}{2}-r \right) \varphi _{{{\it xxx}}}}{24} 
\pm \frac {vy{\varphi^{2} _{{x}}}\varphi _{{{\it xx}}}}{8h}  
+ \frac{ \left( \frac{vy}{2}-r \right) \varphi _{{x}}\varphi _{{{\it xx}}}}{12} 
 \mp \frac {vyh\varphi _{{{\it xx}}}\varphi _{{{\it xxx}}}}{16\varphi _{{x}}}
 \pm \frac {h\kappa\, \left( \theta-vy \right) \varphi _{{{\it xx}}}}{24vy} \\
\\
&& 
\mp \frac {{\varphi^{2} _{{x}}}h \left( \frac{vy}{2}-r \right) ^{2}\varphi _{{{\it xx}}}}{6vy} 
\pm \frac {{\varphi^{2} _{{x}}}h\kappa\, \left( \theta-vy \right) }{24vy} 
+{\rho}^{2}\left[
 \,{\frac {vy{\varphi^{3} _{{x}}}}{3{h}^{2}}}
\mp \,{\frac {vy{\varphi^{2} _{{x}}}\varphi _{{{\it xx}}}}{6h}}
\right]
+ \rho\left[
 \frac{v\varphi _{{x}}\varphi _{{{\it xx}}}}{4}
 + \frac{v{\varphi^{3} _{{x}}}}{12}  \right. \\
 \\
 && \left.
\pm \,{\frac {{\varphi^{4} _{{x}}}\kappa\, \left( \theta-vy \right) }{6hv}}
 - \,{\frac {{\varphi^{3} _{{x}}} \left( \frac{vy}{2}-r \right) }{6y}}
\mp \,{\frac {{\varphi^{2} _{{x}}}h \left( \frac{vy}{2}-r \right) \varphi _{{{\it xx}}}}{6y}} 
\mp \frac{h\varphi _{{{\it xx}}}v{\varphi^{2} _{{x}}}}{24}
\mp \frac{hv{\varphi^{2} _{{{\it xx}}}}}{8}
\pm\,{\frac {{\varphi^{2} _{{x}}}\kappa\, \left( \theta-vy \right) }{6hv}}
\right] ,\\
\\
\hat{K}_{i,j\pm 1} & = &  
 \frac{{\varphi^{3} _{{x}}}\varphi _{{{\it xx}}} \left( \frac{vy}{2}-r \right)}{4} 
 \pm \frac {{\varphi^{3} _{{x}}}h \left( \frac{vy}{2}-r \right) \kappa\, \left( \theta -vy \right) \varphi _{{{\it xx}}}}{4{v}^{2}y} 
 \mp \frac {{\varphi^{2} _{{x}}}h\kappa\, \left( \theta -vy \right) \varphi _{{{\it xxx}}}}{8v} 
 - \frac {5vy{\varphi^{3} _{{x}}}}{12{h}^{2}} 
 + \frac {{\varphi^{3} _{{x}}}v}{12y} \\
\\
&& 
 - \frac {{\varphi^{3} _{{x}}}{\kappa}^{2} \left( \theta -vy \right) ^{2}}{6y{v}^{3}} 
 + \frac {vy\varphi _{{x}}}{12{h}^{2}} 
 \mp \frac {{\varphi^{3} _{{x}}}h\kappa}{12y} 
 \pm \frac {{\varphi^{3} _{{x}}}h{\kappa}^{2} \left( \theta -vy \right) }{12{v}^{2}y} 
 \mp \frac {5\kappa\,{\varphi^{3} _{{x}}} \left( \theta -vy \right) }{12vh} 
 + \frac{vy\varphi _{{x}}{\varphi^{2} _{{{\it xx}}}}}{8} \\
\\
&& 
 + \frac {\kappa\,{\varphi^{3} _{{x}}} \left( \theta -vy \right) }{12vy} 
 + \frac{\kappa\,{\varphi^{3} _{{x}}}}{6} 
 \pm \frac {\varphi _{{x}}\kappa\, \left( \theta -vy \right) }{12vh} 
 \pm \frac {\varphi _{{x}}h{\varphi^{2} _{{{\it xx}}}}\kappa\, \left( \theta-vy \right) }{8v} 
 - \frac{vy{\varphi^{2} _{{x}}}\varphi _{{{\it xxx}}}}{8} \\
 \\
 && 
 +{\rho}^{2} \,{\frac {vy{\varphi ^{3}_{{x}}}}{3{h}^{2}}}
 +\rho\left[
 \frac{v{\varphi^{3} _{{x}}}\varphi _{{{\it xx}}}}{4}
\pm \,{\frac {h{\varphi^{3} _{{x}}}\kappa\, \left( \theta-vy \right) \varphi _{{{\it xx}}}}{4vy}}
\mp \,{\frac {{\varphi^{3} _{{x}}} \left( \frac{vy}{2}-r \right) }{3h}}
 \mp \,{\frac {vy\varphi _{{x}}\varphi _{{{\it xx}}}}{6h}}
\right] ,
\end{array}
\end{eqnarray}
\begin{eqnarray}
\notag \begin{array}{rcl}
\hat{K}_{i+1,j\pm 1} & = & \frac {{\varphi^{4} _{{x}}} \left( \frac{vy}{2}-r \right) }{24h} 
- \frac {vy{\varphi^{2} _{{x}}}\varphi _{{{\it xx}}}}{16h} 
+ \frac { \left( \frac{vy}{2}-r \right) {\varphi^{2} _{{x}}}}{24h} 
+ \frac {vy\varphi _{{{\it xx}}}}{48h} 
- \frac {vy\varphi _{{x}}}{24{h}^{2}} 
- \frac {vy{\varphi^{3} _{{x}}}}{24{h}^{2}} 
\mp \frac {\varphi _{{x}}\kappa\, \left( \theta-vy \right) }{24vh}\\
\\
&& \mp \frac {\kappa\,{\varphi^{3} _{{x}}} \left( \theta-vy \right) }{24vh} 
\pm \frac {\kappa\, \left( \theta -vy \right)  \left( \frac{vy}{2}-r \right) {\varphi^{2} _{{x}}}}{24{v}^{2}y} 
\pm \frac {\kappa\, \left( \theta -vy \right) \varphi _{{{\it xx}}}}{48v} 
\mp \frac { \left( \frac{vy}{2}-r \right) {\varphi^{2} _{{x}}}}{24y} 
 \pm \frac{v{\varphi^{2} _{{x}}}}{48} \\
 \\
 && \pm \frac {{\varphi^{4} _{{x}}}\kappa\, \left( \theta-vy \right)  \left( \frac{vy}{2}-r \right) }{24{v}^{2}y} 
 \mp \frac {\kappa\, \left( \theta-vy \right) {\varphi^{2} _{{x}}}\varphi _{{{\it xx}}}}{16v}  
 +{\rho}^{2}\left[
 \pm  \frac{v{\varphi^{2} _{{x}}}\varphi _{{{\it xx}}}}{8}
+ \,{\frac {vy{\varphi^{2} _{{x}}}\varphi _{{{\it xx}}}}{12h}}
- \,{\frac {vy{\varphi^{3} _{{x}}}}{6{h}^{2}}}
\right] \\
 \\
 &&  +\rho\left[
 \mp \,{\frac {vy{\varphi^{2} _{{x}}}}{4{h}^{2}}}
 \pm \,{\frac {v{\varphi^{2} _{{x}}}}{24y}}
 \pm \,{\frac {{\varphi^{4} _{{x}}}\kappa\, \left( \theta-vy \right) }{24vy}}
 \pm \frac{vy{\varphi^{2} _{{{\it xx}}}}}{48}
 \pm \frac{{\varphi^{4} _{{x}}}\kappa}{24}
 - \,{\frac {{\varphi^{2} _{{x}}}\kappa\, \left( \theta-vy \right) }{12hv}}
 \mp \,{\frac {{\varphi ^{2}_{{x}}}\kappa\, \left( \theta-vy \right) }{24vy}} \right. \\
\\
&& \left.
 \mp \frac{vy\varphi _{{x}}\varphi _{{{\it xxx}}}}{48}
 \pm \,{\frac {{\varphi^{3} _{{x}}} \left( \frac{vy}{2}-r \right) }{6h}}
 \pm \,{\frac {vy\varphi _{{x}}\varphi _{{{\it xx}}}}{12h}}
 \pm \frac{{\varphi^{2} _{{x}}} \left( \frac{vy}{2}-r \right) \varphi _{{{\it xx}}}}{24}
- \,{\frac {{\varphi^{4} _{{x}}}\kappa\, \left( \theta-vy \right) }{12hv}}
 \right] ,
\\
\\
\hat{K}_{i-1,j\pm 1} & = & 
 - \hat{K}_{i+1,j\pm 1}
 - \frac {vy\varphi _{{x}}}{12{h}^{2}} 
- \frac {vy{\varphi^{3} _{{x}}}}{12{h}^{2}} 
\mp \frac {\varphi _{{x}}\kappa\, \left( \theta-vy \right) }{12vh}
\mp \frac {\kappa\,{\varphi^{3} _{{x}}} \left( \theta-vy \right) }{12vh}\\
\\
&&
-  \rho^2 \,{\frac {vy{\varphi^{3} _{{x}}}}{3{h}^{2}}}
 \pm  \rho\left[ \,{\frac {{\varphi^{3} _{{x}}} \left( \frac{vy}{2}-r \right) }{3h}}
 \pm   \,{\frac {vy\varphi _{{x}}\varphi _{{{\it xx}}}}{6h}} \right] 
\end{array}
\end{eqnarray}
and
\begin{eqnarray}\label{Scheme_Heston_Model_Version_4_K_equation_five}
\notag \begin{array}{rcl}
\hat{K}_{i,j} & = & 
 \frac{vy{\varphi^{2} _{{x}}}\varphi _{{{\it xxx}}}}{4} 
 - \frac{{\varphi^{3} _{{x}}}\varphi _{{{\it xx}}} \left( \frac{vy}{2}-r \right)}{2} 
 - \frac{vy\varphi _{{x}}{\varphi^{2} _{{{\it xx}}}}}{4}  
 - \frac {{\varphi^{3} _{{x}}}v}{6y} 
 - \frac {{\varphi^{3} _{{x}}} \kappa\, \left( \theta-vy \right) }{6vy} 
 - \frac{\kappa\,{\varphi^{3} _{{x}}}}{3} 
 - \frac {v\varphi _{{x}}}{6y} 
\\
\\
&& 
 + \frac {{\varphi^{3} _{{x}}}{\kappa}^{2} \left( \theta-vy \right) ^{2}}{3y{v}^{3}} 
 + \frac {5vy\varphi _{{x}}}{6{h}^{2}}  
 + \frac {5vy{\varphi^{3} _{{x}}}}{6{h}^{2}} 
 - \frac{ \left( \frac{vy}{2}-r \right) \varphi _{{x}}\varphi _{{{\it xx}}}}{6} 
 - \frac{vy\varphi _{{{\it xxx}}}}{12} 
 + \frac {{\varphi^{3} _{{x}}} \left( \frac{vy}{2}-r \right) ^{2}}{3vy} \\
\\
&& 
 + \frac {\varphi _{{x}}\kappa\, \left( \theta-vy \right) }{6vy} 
 - {\rho}^{2}   \,{\frac {2vy{\varphi^{3} _{{x}}}}{3{h}^{2}}} 
 +\rho\left[
- \frac{v\varphi _{{x}}\varphi _{{{\it xx}}}}{2}
+ \,{\frac {{\varphi^{3} _{{x}}} \left( 1/2\,vy-r \right) }{3y}}
- \frac{v{\varphi^{3} _{{x}}}}{6}
- \frac{v{\varphi^{3} _{{x}}}\varphi _{{{\it xx}}}}{2}
\right]
\end{array}
\end{eqnarray}
where $\hat{K}_{i,j}$ is the coefficient of $U_{i,j}(\tau)$. Defining $\hat{M}_{i,j}$ as the coefficient of $\partial_{\tau}U_{i,j}(\tau)$ we get
\begin{eqnarray}
\notag \begin{array}{rcl}
\hat{M}_{i+1,j\pm 1} & = & \hat{M}_{i-1,j\mp 1} = \pm \,\rho\frac{{\varphi^{4} _{{x}}}}{24} ,
\\
\\
\hat{M}_{i \pm 1,j} & = & 
 \frac{{\varphi^{3} _{{x}}}}{12}
 \mp  \,{\frac {{\varphi^{4} _{{x}}}h \left( \frac{vy}{2}-r \right) }{12vy}}
 \pm \frac{{\varphi^{2} _{{x}}}h\varphi _{{{\it xx}}}}{8}
 \mp \rho\,{\frac {{\varphi^{4} _{{x}}}h}{12y}} ,
\\
\\
\hat{M}_{i,j \pm 1} & = & 
 \frac{{\varphi^{3} _{{x}}}}{12}
 \pm \,{\frac {{\varphi^{3} _{{x}}}h\kappa\, \left( \theta-vy \right) }{12{v}^{2}y}}
 \mp \,{\frac {{\varphi^{3} _{{x}}}h}{12y}} \text{ and}
\\
\\
\hat{M}_{i,j} & = & 
 \frac{2{\varphi^{3} _{{x}}}}{3}
- \,{\frac {{\varphi^{3} _{{x}}}\varphi _{{{\it xx}}}{h}^{2} \left( \frac{vy}{2}-r \right) }{2vy}}
- \frac{\varphi _{{x}}{\varphi^{2} _{{{\it xx}}}}{h}^{2}}{4}
+ \frac{{\varphi^{2} _{{x}}}{h}^{2}\varphi _{{{\it xxx}}}}{4}
- \rho \,{\frac {{\varphi^{3} _{{x}}}\varphi _{{{\it xx}}}{h}^{2}}{2y}} .
\end{array}
\end{eqnarray}
Using these coefficients instead of the ones given in \eqref{Scheme_Heston_Model_Version_3_K_equation_one} to \eqref{Scheme_Heston_Model_Version_3_M1_to_M9} in the derivation in Section~\ref{sec:HOC_schemes_for_parabolic_problems} for the interior of the grid $G$ as well as the boundaries $y_{\text{min}}$ and $y_{\text{max}}$ yields the Version 4 scheme.
